\documentclass[aps, prb, reprint, showpacs, superscriptaddress]{revtex4-1}

\usepackage{amsfonts}
\usepackage{amsmath}
\usepackage{amssymb, bm}
\usepackage{graphicx, epstopdf, color, braket}
\usepackage[dvipsnames]{xcolor}
\usepackage{soul}
\usepackage{cleveref, appendix,tikz}
\usetikzlibrary{patterns,arrows,decorations.pathreplacing}
\usetikzlibrary{plotmarks}

\begin{document}

\title{Valleytronics in merging Dirac cones: all-electric-controlled valley filter, valve and universal reversible logic gate}

\author{Yee Sin Ang}
\email{Corresponding author. yeesin\_ang@sutd.edu.sg}
\affiliation{SUTD-MIT International Design Center, Singapore University of Technology and Design, Singapore 487372}

\author{Shengyuan A. Yang}
\affiliation{SUTD-MIT International Design Center, Singapore University of Technology and Design, Singapore 487372}

\author{C. Zhang}
\affiliation{School of Physics, University of Wollongong, NSW 2522, Australia}

\author{Zhongshui Ma}
\affiliation{School of Physics, Peking University, Beijing 100871, China}
\affiliation{Collaborative Innovation Center of Quantum Matter, Beijing 100871, China}

\author{L. K. Ang}
\email{Corresponding author. ricky\_ang@sutd.edu.sg}
\affiliation{SUTD-MIT International Design Center, Singapore University of Technology and Design, Singapore 487372}

\begin{abstract}
	
	Despite much anticipation of valleytronics as a candidate to replace the ageing CMOS-based information processing, its progress is severely hindered by the lack of practical ways to manipulate valley polarization all-electrically in an electrostatic setting. 
	Here we propose a class of all-electric-controlled valley filter, valve and logic gate based on the valley-contrasting transport in a merging Dirac cones system.
	The central mechanism of these devices lies on the pseudospin-assisted quantum tunneling which effectively quenches the transport of one valley when its pseudospin configuration mismatches that of a gate-controlled scattering region.
	The valley polarization can be abruptly switched into different states and remains stable over semi-infinite gate-voltage windows. Colossal tunneling valley-pseudo-magnetoresistance ratio of over 10,000\% can be achieved in a valley-valve setup. 
	We further propose a valleytronic-based logic gate capable of covering all 16 types of two-input Boolean logics.
	Remarkably, the valley degree of freedom can be harnessed to resurrect logical-reversibility in two-input universal Boolean gate.
	The (2+1) polarization states -- two distinct valleys plus a null polarization -- re-establish one-to-one input-to-output mapping, a crucial requirement for logical-reversibility, and significantly reduce the complexity of reversible circuits due to the built-in nature of valley degree of freedom.
	Our results suggest that the synergy of valleytronics and digital logics may provide new paradigms for valleytronic-based information processing and reversible computing.

\end{abstract}

\maketitle

\section{Introduction}

Valleytronics is an emerging device concept \cite{rycerz, xiao1, yao} based on the manipulation of \emph{valley degree of freedom} in certain condensed matter systems such as semiconductor quantum well \cite{semicond}, silicon \cite{silicon}, bismuth \cite{bismuth}, diamond \cite{diamond}, carbon nanotube \cite{cnt}, graphene \cite{graphene}, Dirac semimetal \cite{ds} and transition metal dichalcogenide (TMD) monolayer \cite{mos2}.
In these materials, electrons can populate multiple low energy states that are well separated in momentum-space, known as valley. 
The electron's `valley address', or the valley degree of freedom, provides an additional quantum index which can be harnessed for new paradigm of classical and quantum information processing \cite{vb}. Valleytronics, alongside with spintronics \cite{spin}, photonics and plasmonics \cite{photon}, has been proposed as a candidate system to replace the aging CMOS technology \cite{itrs2015}.

\begin{figure}[t]
	\includegraphics[scale=.58]{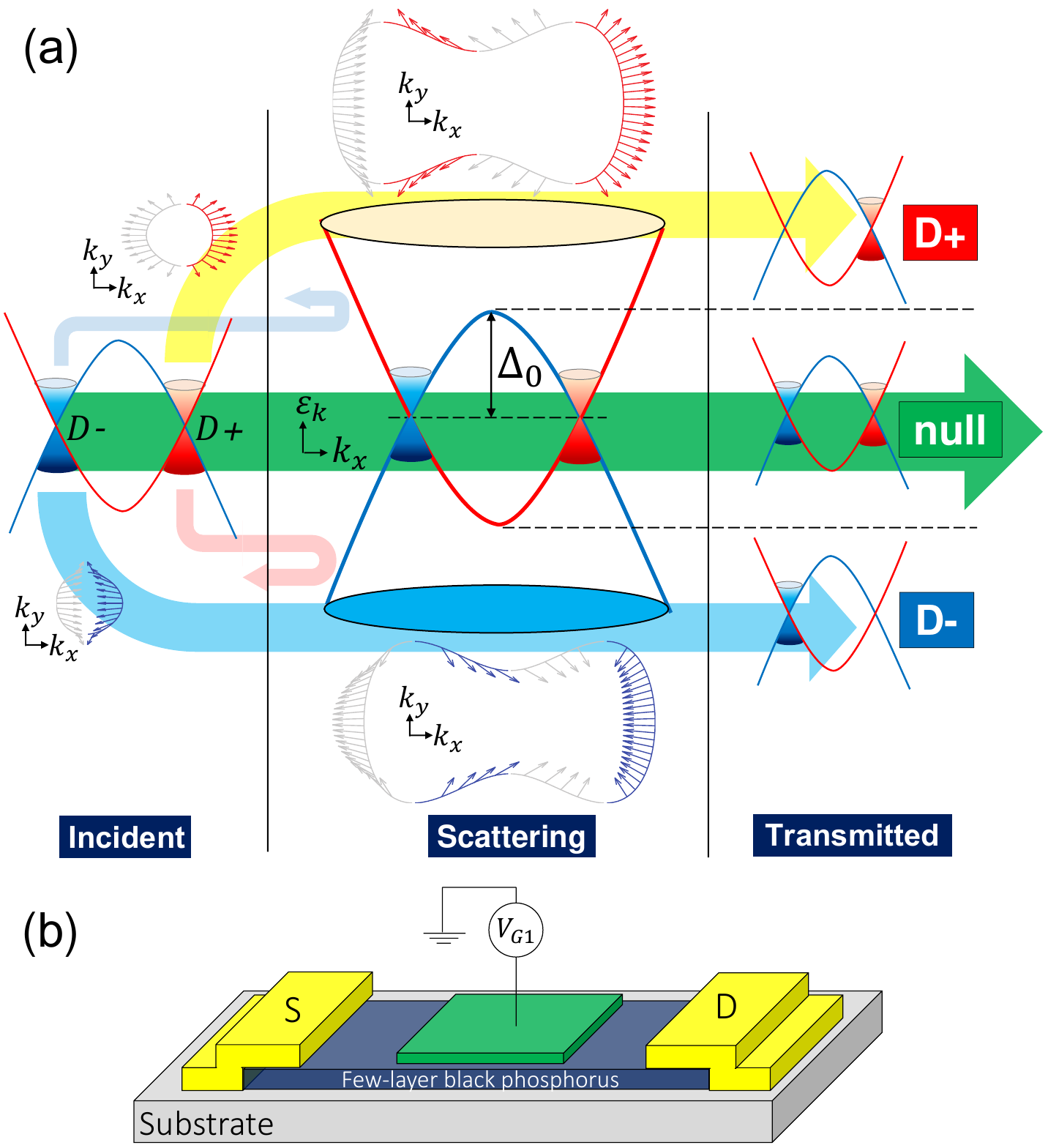}
	\caption{Concepts of all-electric-controlled valley filter. (a) Band diagram of pseudospin-assisted valley-contrasting quantum tunneling structure in a two-dimensional merging Dirac cone system. The matching and mismatching of pseudospin configuration between the incident and the scattering regions allow the valley polarization of the transmitted current to be all-electrically tuned. The iso-energy contours are shown alongside with the pseudospin vectors of forward (towards-right) propagating states. Backward (towards-left) propagating states are grayed-out. (b) Schematic drawing of a valley filter based on a transistor setup.
	}
\end{figure}

Despite recent success in optical manipulation of valley in TMDs \cite{mos2, abergel}, the experimental progress of generating valley polarization via d.c. approach remains stagnant due to the lack of practical \emph{valley filter} -- a device that produces valley-polarized current. 
In general, valley filters can be classified into two types: (i) gauge-field-based (GF); and (ii) electrostatic-field-based (EF).
GF filter \cite{fujita, zhai, low, zhai2, zhai3, zhai4, zhai5, wu, jiang, cai, pratley, grujic, cavalcante, settnes, settnes2, sekera} utilizes an external magnetic field and/or a pseudo-magnetic field induced by mechanically straining the crystal \cite{fogler} to break the valley transport symmetry whereas EF filter mostly relies on energy filtering in properly designed nanostructures \cite{rycerz, ezawa,xu} or by forming 1D topological edge state in a domain wall \cite{martin, qiao, ezawa2, fzhang, qiao2, pan, anglin, ju, li}.
In terms of building compact valleytronic device, EF filter is more advantageous than GF filter as the electrical output of an EF filter is intrinsically compatible with its electric-based controlling knob for valley polarization. This is in contrast to GF filter in which cascading multiple filters would require the formidable tasks of on-chip electricity-to-magnetic or electricity-to-strain conversions to be tamed.

Albeit the practical usefulness of EF valley filter, only a small subset of filters are capable of all-electric-control due to the difficulty in breaking valley transport symmetry solely via electrostatic field. 
Moreover, these filters are severely plagued by stringent conditions such as the need of high-precision structural control of nanostructures or ultra-low operating temperature to prevent bulk current from flooding the subtle valley signal carried by 1D topological edge state \cite{ju,li}.
Thus far, the search for an easy-to-implement, all-electric-field-controlled valley filter remains an ongoing challenge.
Beyond valley filters, \emph{valley beam splitter}, operating via an electron-optics approach \cite{elec_opt}, has been explored as an alternative building block of valleytronics \cite{beam_split}. 

In this work, we propose a class of all-electric-controlled valley filtering based on the \emph{pseudospin-assisted valley-contrasting quantum tunneling} in quasi two-dimensional system with merging Dirac cones (2MDS) which can be created in a wide class of systems including honeycomb lattice of cold atoms \cite{taruell}, graphene \cite{montambaux, delplace}, few-layer black phosphorus \cite{BP_SD, baik, BP_E, BP_S, BP_P, deng, BP_L}, Weyl semimetal \cite{ds_MDS} and antimonene -- single layer of antimony \cite{sb}. 
The valley polarization is fully gate-controlled and is robust against gate voltage fluctuation.
By arranging two filters into \emph{valley valve}, this pseudospin-assisted filtering effect can produce a colossal \emph{valley-pseudo-magnetoresistance} ratio of well over 10,000\%. This colossal ratio dwarves the tunneling magnetoresistance in conventional magnetic tunnel junctions \cite{TMR} and the pseudo-magnetoresistance in graphene-based pseudospin valve \cite{san-jose}, and is on par with the colossal tunneling electroresistance effect in state-of-art ferroelectric tunnel junctions \cite{TER} .
We further propose a concept of \emph{valleytronic logic gates} which encompasses all 16 types of two-input Boolean logics.

More remarkably, the valley degree of freedom, which manifests macroscopically in a (2+1) fashion -- two valley polarizations plus a null polarization state, can be harnessed as a built-in `valley-pigeonhole' for input information storage. This offers a unique possibility of implementing \emph{logically-reversible universal Boolean gate}, a precursor of dissipationless classical reversible computer \cite{feynman}, in a valleytronic system. 
Our results reveal a concrete architecture of valleytronic-based digital information processing. 
The synergy of valleytronic and Boolean logic may provide a viable new route towards reversible computation which is ultimately required to minimize waste heat generation in classical computer.

\begin{figure}[t]
	\includegraphics[scale=.25]{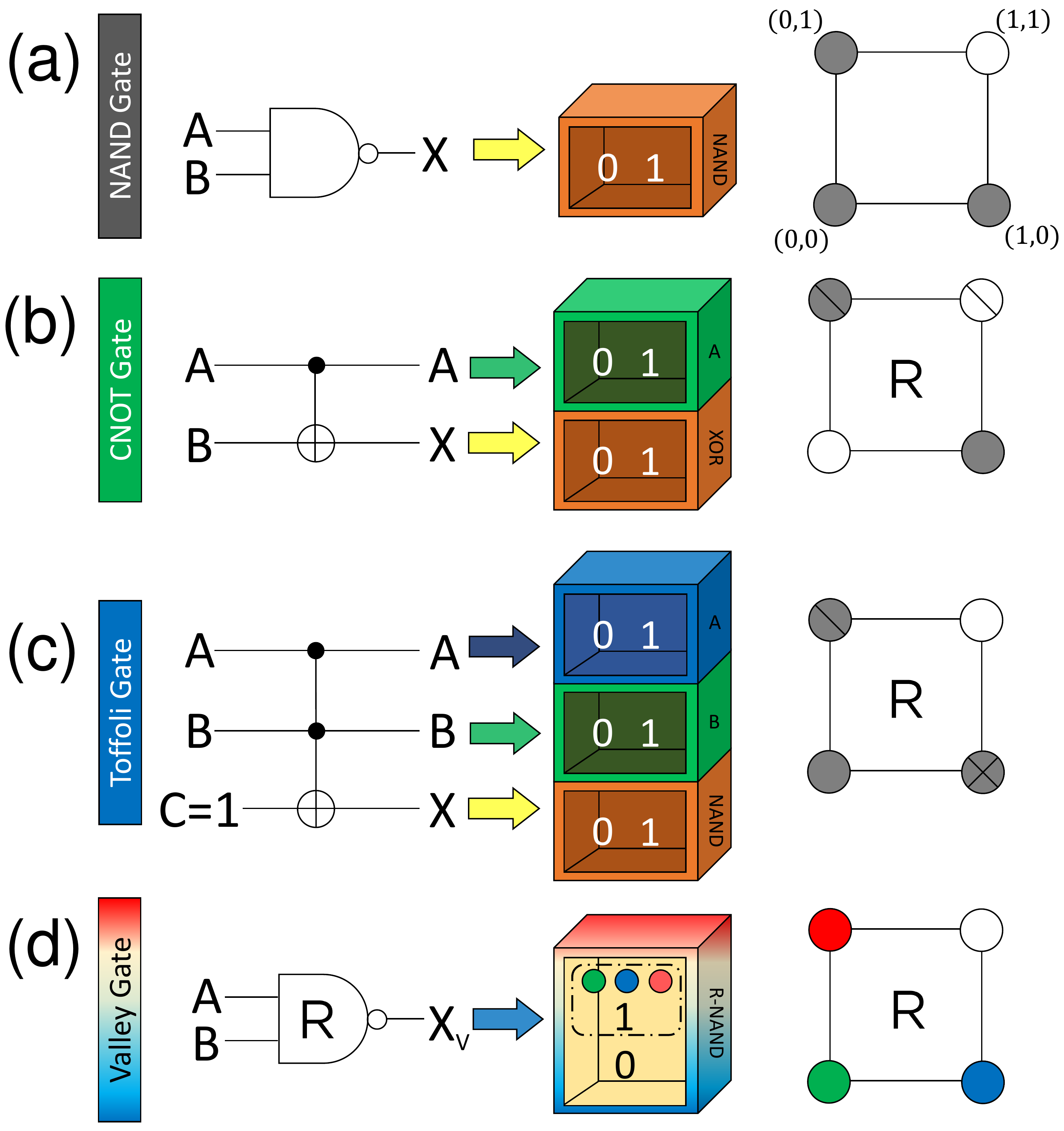}
	\caption{Reversibility and Boolean loop representations of logic gates. The four vertices of the Boolean loop represents various $(A,B)$ input configuration. Filled (emptied) node denotes `1' (`0') output. Letter `R' emphasizes logical-reversibility. (a) Traditional irreversible NAND. (b) Reversible CNOT. (c) Reversible NAND based on Toffoli gate by holding $C = 1$.
		(d) Valleytronic-based reversible NAND.}
\end{figure}

\subsection{Concept of pseudospin-assisted valley filter}

The central operating mechanism of the valley filter lies on the \emph{pseudospin-assisted valley-contrasting quantum tunneling} which effectively quenches the transport of one valley when its pseudospin mismatches that of a gate-controlled scattering region [Fig. 1]. 
In the following, we shall use 2MDS in few-layer black phosphorus as a model structure to illustrate the valley filtering effect. It is proposed that the bandgap of two-dimensional black phosphorus can be engineered via perpendicular electric field, surface doping, pressure or laser irradiation \cite{BP_SD, baik, BP_E, BP_S, deng, BP_P, BP_L}. 
Bandgap tuning \cite{deng}, bandgap closure and band inversion \cite{BP_SD} of few-layer black phosphorus have been realized in recent experiments. 
The band inversion regime offers a particularly interesting platform for valleytronic applications due to the emergence of two well-separated Dirac cones. 
In the low energy regime $|\varepsilon_k| < \Delta_0$ where $\Delta_0$ is a bandgap parameter, the energy spectrum is composed of two Dirac cones, denoted as $D+$ and $D-$ valleys, with opposite \emph{chirality} -- the pseudospin vector near the Dirac point is locked to the quasiparticle wavevector, and its winding configuration is opposite between the two valleys [Fig. 1(a)]. 
For $|\varepsilon_k| > \Delta_0$, the two valleys merge into a single Fermi surface. 
In this case, although the valleys are no longer well-defined, the pseudospin of forward propagating states still orientates in a fashion that resembles the pseudospin chirality of $D+$ valley for $\varepsilon_k >\Delta_0$ and of $D-$ valley for $ \varepsilon_k < -\Delta_0$. 
This creates two well-separated energy windows in which only electrons from a valley of matching chirality are favorably transported due to the conservation of pseudospin \cite{klein}.
In a device sense, such unusual band topology can be harnessed to construct a valley filter via a `source-channel-drain' transistor setup [Fig. 1(b)]. 
By gate-tuning the Fermi energy, $\varepsilon_F$, of the channel between the windows of $\varepsilon_F > \Delta_0$, $\varepsilon_F< - \Delta_0$ and $|\varepsilon_F| < \Delta_0$, the valley polarizations of the transmitted electrical current can be switched between $D+$, $D-$ valleys, and null polarization, respectively. 
Apart from being all-electric-controllable, the valley polarization remains remarkably stable over the \emph{semi-infinite} energy windows of $\varepsilon_F > \Delta_0$ and $\varepsilon_F < -\Delta_0$ which can be particularly useful for device applications. The merging transition of 2MDS in few-layer black phosphorus has been experimentally demonstrated to occur with a sizable energy window of $ \approx - 200$ meV \cite{BP_SD}. Such wide energy window may prove a suitable platform for the manipulation of valley in 2MDS provided that the stability and device-fabrication issues of few-layer black phosphorus can be overcome.

\subsection{Concept of reversible valleytronic gate}

The presence of valley degree of freedom adds a new dimension to the Boolean operation in terms of logical-reversibility.
The reversibility of Boolean logical operation is illustrated in Figs. 2(a)-(d) by using \emph{Boolean loop} -- a graphical representation of Karnaugh map \cite{karnaugh} [see Fig. 1(c)] -- in which the four vertices represent all $2^2$ possible input configurations (the two inputs are represented by $(A,B)$ where $A=0,1$ and $B = 0,1$) and the output state is encoded as followed: empty and filled node denote `0' and `1' output state, respectively.  
Traditional Boolean gate, such as NAND gate [Fig. 2(a)], is \emph{logically-irreversible} due to the simultaneous presence of multiple filled-nodes, i.e. the input is mapped into output via a many-to-one fashion. 
As the outputs cannot be fully unambiguously reversed back to their corresponding input, part of the input information is inevitably lost.
Such logical-irreversibility has a profound practical implication -- the energy efficiency of Boolean-based computer is ultimately capped at the \emph{Landauer's limit}, an irreducible waste heat generation of $k_BT \ln 2$ per bit of information erased \cite{landauer}. 

One potential route to break Landauer's limit is envisaged to be provided by \emph{reversible computation} which processes information reversibly \cite{feynman}. 
Its precursor, \emph{universal logic gate}, has been actively searched for since early 1970s' \cite{toffoli, fredkin, hardy}. 
Controlled-NOT gate -- an XOR gate supplemented by an extra output bit identical to one of its input -- represents a classic reversible gate \cite{feynman2}. 
The supplementary bit serves as an information pigeonhole to hosts two distinct `colors' [denoted by slashed nodes in Fig. 2(b)] that can be used to fully remove the double-ambiguity of XOR operation.
Universal reversible gate, such as Toffoli gate which provides reversible NAND operations on $(A,B)$ by holding `C' input in `1' state, requires two supplementary bits to generate $2^2$ distinct `colors' in order to fully remove the \emph{triple-ambiguity} in the output state. Thus, three out of the $2^2$ `colors' [denoted by filled, slashed and crossed nodes in Fig. 2(c)] are required to simultaneously preserve both universality and logical-reversibility of two-input Boolean gate. 

In contrast, the valleytronic-based reversible logic gate proposed in this work operates in a fundamentally different way. The output current produced by the valleytronic system is additionally dressed by (2+1) distinct 'valley-colors', i.e. two possible states of valley polarizations plus a null polarization state. 
These built-in 'valley-colors' [denoted, respectively, by red, blue and green nodes in Fig. 2(d)] can be readily utilized to establish one-to-one mapping between input and output states of a NAND gate. Logical-reversibility and universality can thus be simultaneously achieved without the need of adding supplementary bits.
More importantly, this allows the valleytronic-based reversible logic gate to retain the simple two-input architecture of conventional Boolean gate and is in stark contrast to the more complicated three-input architecture of Toffoli and Fredkin gates \cite{toffoli, fredkin}.

\section{Model}

\begin{figure} [t]
	\includegraphics[scale= .85]{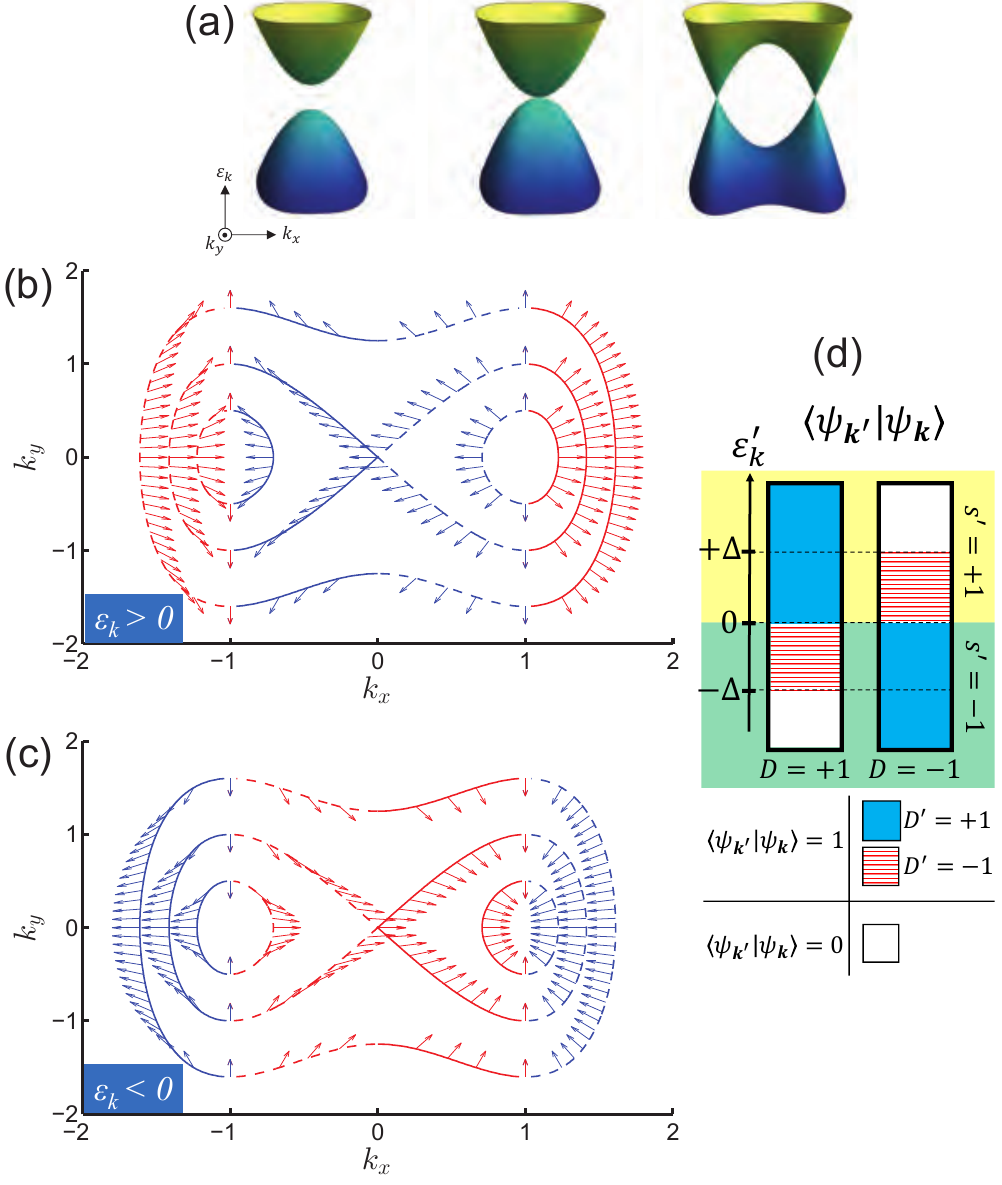}
	\caption{Band topology, pseudospin configuration and valley-contrasting transport in merging Dirac cone system. (a) Topological transition of the band structure for $\Delta>0$, $\Delta =0$; and $\Delta<0$. (b) Pseudospin texture for $\varepsilon_{\mathbf{k}} > 0$ and for (c) $\varepsilon_{\mathbf{k}} > 0$. Forward and backward propagation is denoted by solid and dashed contour lines, respectively. Blue and red colors denote quasiparticle states with $S_x < 0$ and $S_x>0$, respectively. For (b) and (c), $\Delta = -1$ is used. The innermost, intermediate and outermost contour lines represent $\varepsilon_{\mathbf{k}} = (0.5, 1, 1.6)$ and $\varepsilon_{\mathbf{k}} = (-0.5, -1, -1.6)$, respectively in (b) and (c). (d) Diagram of $\braket{\psi_{\mathbf{k}',s'}|\psi_{\mathbf{k},s' }}$ for $s=+1$ (left and right bars corresponds to $\mathcal{D}' = \pm 1$). Empty region denotes $\braket{\psi_{\mathbf{k}',s'}|\psi_{\mathbf{k},s' } } = 0$. Filled and horizontally-striped region denote $\braket{\psi_{\mathbf{k}',s'}|\psi_{\mathbf{k},s' } } = 1$ with $\mathcal{D}' = \pm1$, respectively.} 
\end{figure}

The merging transition of the two Dirac cones in 2MDS can be modeled by an effective Hamiltonian proposed by Montambaux et al \cite{montambaux},
\begin{equation}
\hat{\mathcal{H}}_\mathbf{K} = 
\left(\frac{\hbar^2K_x^2}{2m^*} + \tilde{\Delta} \right)\sigma_x + v_F K_y\sigma_y,
\end{equation}
where $\mathbf{K} = (K_x, K_y)$ is the wavevector, $m^*$ is the effective mass along $K_x$-direction, $v_F$ is the Fermi velocity along $K_y$-direction, $\bm{\sigma} = (\sigma_x,\sigma_y)$ is the Pauli pseudospin matrix and $\tilde{\Delta}$ is a bandgap parameter. For simplicity, we transform Eq. (1) into a dimensionless form via the following definitions: $\mathbf{k} \equiv \mathbf{K}/ k_0$ and $\hat{\mathcal{H}}_{\mathbf{k}} \equiv \hat{\mathcal{H}}_{\mathbf{K}} / \varepsilon_0$ where $k_0 = 2m^*v_F/\hbar$ and $\varepsilon_0 = 2m^*v_F^2$ are defined as the characteristic wavevector and energy respectively. 
The dimensionless Hamiltonian takes the form of $\hat{\mathcal{H}}_{\mathbf{k}} = \left( k_x^2 + \Delta \right) \sigma_x + k_y \sigma_y$, where $\Delta \equiv \tilde{\Delta} / \varepsilon_0$.
The energy dispersion is $\varepsilon_{\mathbf{k}} = s \sqrt{\left(k_x^2+\Delta\right)^2 + k_y^2}$ where $s = \pm 1$ denotes conduction and valence band, respectively. Such dispersion exhibits a `semi-Dirac' behavior, i.e. $\varepsilon_\mathbf{k}$ along $k_x$-axis and $k_y$-axis exhibits non-relativistic parabolic and ultra-relativistic linear dispersion, respectively \cite{semidirac}. 
The band topology is crucially determined by the sign of $\Delta$ [Fig. 3(a)]. For $\Delta > 0$, the system is a band insulator. The bandgap gradually closes as $\Delta \to 0$. Band inversion, accompanied by the emergence of two Dirac cones situated along the $k_x$-axis at $\mathbf{k} = (\pm \sqrt{\left|\Delta \right|}, 0)$, occurs for $\Delta < 0$. In this case, the Fermi surface is made up of two distinct Dirac pockets of opposite chirality in the low energy regime of $|\varepsilon_{\mathbf{k}}| < |\Delta|$ (see Appendix A). The Dirac pockets merge into a single Fermi surface for $|\varepsilon_{\mathbf{k}}| > |\Delta|$. 

We now focus on the $x$-directional transport with $\Delta <0$ where the electron transport exhibits dramatic valley filtering effect.
In the presence of a scattering potential $U(x)$, $k_x$ is replaced by $k_x \to -i \partial/\partial_x$. The eigenstate can be solved from the Schr\"odinger equation, $\hat{\mathcal{H}}_\mathbf{k} \psi = \varepsilon_\mathbf{k} \psi$, to yield an eigenstate of $\psi^{(\lambda\eta)}(\varepsilon_{\mathbf{k}}, k_y) = [ 1, \left(\lambda \sqrt{\varepsilon_{\mathbf{k}}^2 - k_y^2} + ik_y \right)/  \varepsilon_{\mathbf{k}} ]^{\mathcal{T}} e^{ik_x^{(\lambda\eta)}x}$, where $\mathcal{T}$ denotes transpose, $k_x^{(\lambda\eta)} = \lambda \sqrt{\eta \left( \varepsilon_{\mathbf{k}}^2 - k_y^2 \right)^{1/2} -\Delta }$, $\lambda = \pm1$ and $\eta = \pm1$ label the four eigenstates. 
Figs. 3(b) and (c) illustrate the pseudospin texture, given by $\mathbf{S} = (S_x, S_y) = \psi^{(\lambda\eta) \dagger} \bm{\sigma} \psi^{(\lambda\eta)}$, along several iso-energy contours with $\varepsilon_{\mathbf{k}}>0$ and $\varepsilon_{\mathbf{k}}<0$, respectively. Two unusual features are observed.
First, the $S_y$-component, $S_y = k_y / \varepsilon_{\mathbf{k}}$, is identical between the two valleys while the $S_x$-component, $S_x = (k_x^{(\lambda\eta)2} + \Delta ) / \varepsilon_{\mathbf{k}}$ is strongly $k_x$-dependent which leads to valley-contrasting transport occurs along the $x$-direction.
Second, the two valleys exhibit opposite pseudospin winding configurations and their forward propagating states carries $S_x > 0$ and $S_x<0$, respectively, for $D+$ and $D-$ valleys [denoted by red and blue arrows in Figs. 3(b) and (c)]. More importantly, once the two Dirac pockets merge into a single Fermi surface when $\left|\varepsilon_k\right| > \left| \Delta \right|$, the forward propagation becomes dominated by $S_x > 0$ for $\varepsilon_{\mathbf{k}} > \left| \Delta \right|$ and $S_x< 0$ for $\varepsilon_{\mathbf{k}} < - \left| \Delta \right|$. Thus, the transmission of $D-$ and $D+$ valley states becomes strongly preferred within the semi-infinite energy windows of $\varepsilon_{\mathbf{k}} > \left|\Delta\right|$ and $\varepsilon_{\mathbf{k}} < - \left| \Delta \right|$, respectively. 

This valley contrasting transport can be further illustrated via an 1D scattering model (i.e. $k_y = 0$) in which an initial state, $\ket{\psi_{\mathbf{k,s}}}$, is scattered into a final state, $\ket{\psi_{\mathbf{k'},s'}}$, by a pseudospin non-flipping potential $\hat{V}$, as characterized by $\braket{\psi_{\mathbf{k'},s'}|\hat{V}|\psi_{\mathbf{k},s} } = V \braket{\psi_{\mathbf{k'},s'}|\psi_{\mathbf{k},s} }$ with $\mathbf{k}' \neq \mathbf{k}$ and $V$ is a $\mathbf{k}$-independent potential strength. 
The transition amplitude can be obtained as $\braket{\psi_{\mathbf{k'}}|\psi_\mathbf{k}} = (1 + ss'\mathcal{D}\mathcal{D}')/2$ where $\mathcal{D} = \text{sign}(k_x^2 + \Delta)$ and $\mathcal{D}' = \text{sign}(k_x^{'2} + \Delta)$ represent the valleys of $\ket{\psi_{\mathbf{k,s}}}$ and $\ket{\psi_{\mathbf{k',s'}}}$, respectively. 
Note that the valleys are labeled by $\mathcal{D}$ and $\mathcal{D}'$ as followed. For $(s, s') = + 1$, the $D\pm $ valley is represented by $(\mathcal{D}, \mathcal{D}') = \pm 1$. For $(s,s' ) = -1$, the $D\mp$ valley is represented by $(\mathcal{D}, \mathcal{D}') = \pm 1$.
For simplicity, we only consider an initial state with $s=1$. 
For a final state with $\varepsilon_{\mathbf{k}'} < |\Delta|$, $\mathcal{D}'$ can be either $\pm1$ due to the presence of two distinct Dirac branches.
In this case, $\braket{\psi_{\mathbf{k'}}|\psi_\mathbf{k}} = 1$ occurs when the condition, $s'\mathcal{D}\mathcal{D}' = 1$, is fulfilled. 
Such condition can be translated as followed: the scattering of $\ket{\psi_{\mathbf{k,s}}}$ into $\ket{\psi_{\mathbf{k'},s'}}$ is allowed either via intraband ($s'=+1$, $\mathcal{D} = \mathcal{D}'$) or via interband ($s'=-1$, $\mathcal{D} = -\mathcal{D}'$) pathway for any valley index (i.e. $\mathcal{D}\pm1$) of the initial state. This corresponds to an `all-pass' scenario where both valleys can simultaneously exhibit unity transition amplitude. 
In contrast, for a final state with $\varepsilon_{\mathbf{k}'} > |\Delta|$, the convergence of two Dirac branches leads to the only possibility of $\mathcal{D}' = +1$. This results in a more stringent condition of $s'\mathcal{D} = 1$ for $\braket{\psi_{\mathbf{k'},s'}|\psi_{\mathbf{k},s} } = 1$ that corresponds to an exclusively one-valley scattering process of either $\mathcal{D} = +1$ via intraband ($s'=1$) pathway or $\mathcal{D} = -1$ via interband ($s'= -1$) pathway. Various possibility of $\braket{\psi_{\mathbf{k}',s'}|\psi_{\mathbf{k},s' } }$ is summarized in Fig. 3(d). Here, the quasiparticle scattering can be valley-selectively controlled by switching the final state band index, $s'$. This valley-selective scattering effect forms the central operating mechanism of the valleytronic devices proposed in this work. 

\begin{figure*} [t]
	\includegraphics[scale= 1.55]{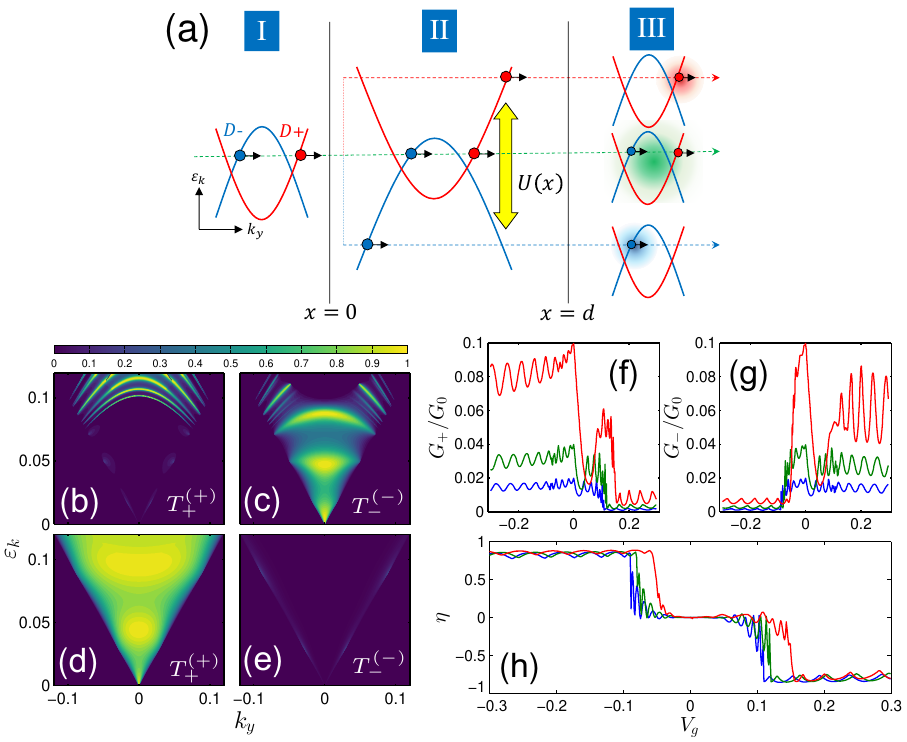}
	\caption{Operation of valley filter. (a) Band digram of the valley filter. $\Delta$ is constant throughout Regimes I, II and III. The band alignment of Regime II is gate-tunable via $U(0<x<d)$. Transmission probabilities, $T^{(+)}_+$, for (b) potential barrier and for (d) potential well. (c) and (e): same as (b) and (d) for $T^{(-)}_-$. Valley-polarized conductance in (g) $D+$ and (g) $D-$ channel. Blue, green and red curves correspond to $\varepsilon_F/ \left|\Delta\right| = (0.1, 0.2, 0.5)$, respectively. (h) Valley polarization efficiency, $\eta$. $\Delta = -0.1$ and $d = 75$ are used.}
\end{figure*}

\begin{figure*} [t]
	\includegraphics[scale= 1.2 ]{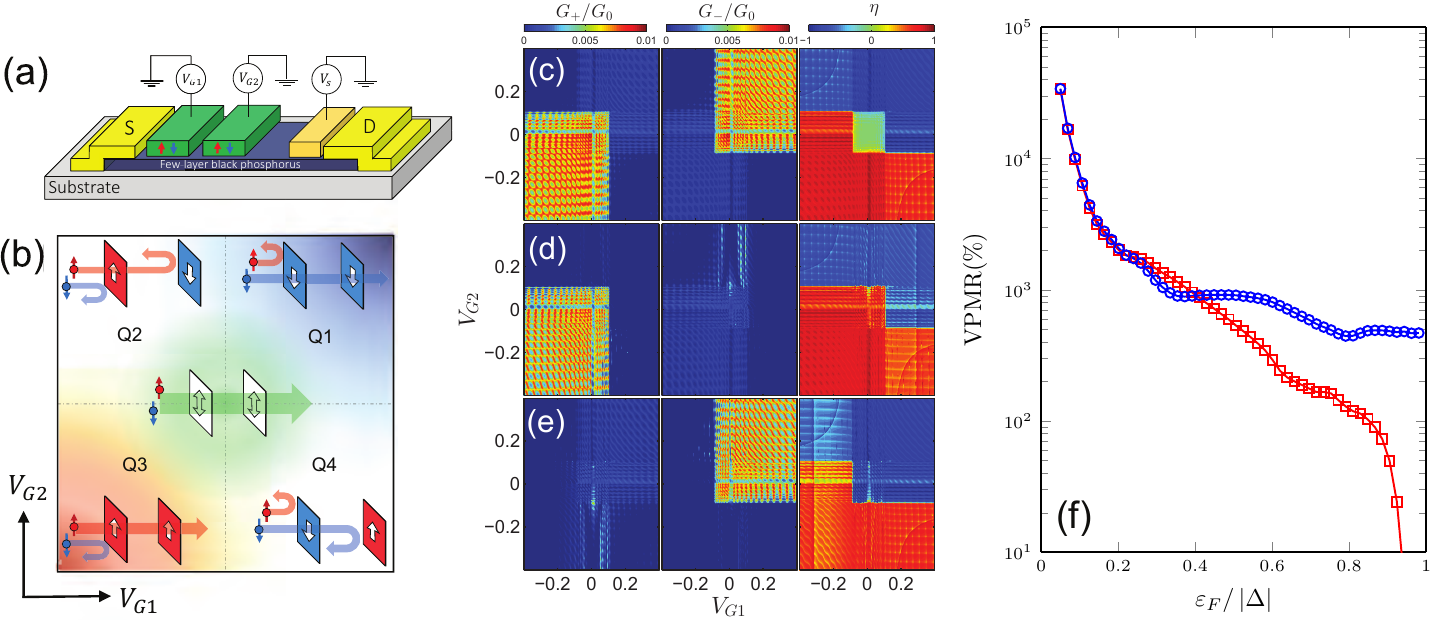}
	\caption{Operation of valley valve and colossal valley-pseudo-magnetoresistance effect. (a) Schematic drawing of valley valve. (b) Parallel and anti-parallel configurations and their respective quadrant in the $V_{G1}$-$V_{G2}$ space. (c) Valley-dependent conductance $G_{\pm}$ and the corresponding $\eta$. (d) and (e): same as (c) but operated in $D+$ and $D-$ mode, respectively, via $V_s = -0.2$ and $V_s = 0.2$. (f) VPMR ratio of Q1$\leftrightarrows$Q2 (denoted by red squares) and Q3$\leftrightarrows$Q2 (denoted by blue circles) is shown. The VPMR ratios of Q1$\leftrightarrows$Q4 and Q3$\leftrightarrows$Q4 is approximately identical to this plot and hence are not shown. The following parameters are used: $\Delta = -0.1$, $\varepsilon_F/\left|\Delta\right| = 0.1 $, the barrier width corresponding to $V_{G1, 2, s}$ is set to $d_{\text{gate}} = 75$ and are separated by $d_{\text{inter}} = 20$.}
\end{figure*}

\section{Valleytronic trio: filter, valve and reversible logic gate}

In this section, we show that the highly non-trivial band topology and the pseudospin texture in 2MDS can be harnessed to create a trio of all-electric valleytronic devices: valley filter, valve and reversible logic gate (see Appendices B and C for details of device modeling). In the following, the device modeling is performed using band structure parameters of few-layer black phosphorus with merging Dirac cones calculated from first-principle by Baik et al \cite{baik}, i.e. $\varepsilon_0 \approx 1.3$ eV and $k_0 \approx 1.1$ nm$^{-1}$ \cite{parameter}. We model the valleytronic devices using a Landauer's ballistic transport formalism \cite{landauer2} for 2D nanostructures. The ballistic transport picture has been widely used in the modeling of valley filtering effect in nanostructures \cite{rycerz, fujita, zhai, low, zhai2, zhai3, zhai4, zhai5, wu, jiang, cai, pratley, grujic, cavalcante, settnes, sekera,xu}. In realistic device, the inevitable presence of impurities, defects and many-body effects can quantitatively change the results, but the valley filtering effect shall qualitatively remain robust as recently demonstrated for the case of strained graphene \cite{settnes2}.

\subsection{Chiral valley filtering effect}

As a proof-of-concept, we first demonstrate the \emph{pseudospin-assisted} valley-selective quantum tunneling, with band diagram shown in Fig. 4(a), by calculating the same-valley transmission probability $T_\pm^{(\pm)}$ as a function of incident energy, $\varepsilon_{\mathbf{k}}$, and gate voltage, $V_g$ [Figs. 4(b)-(c)]. The $\mathcal{D} \to \mathcal{D}'$ transmission probabilities is denoted as $T_{\mathcal{D'}}^{(\mathcal{D})}$ where $\mathcal{D}, \mathcal{D}' = \pm1$ represents different valley states.
The inter-valley scattering effect is intrinsically included in this model. 
Its scattering probabilities, $T_\mp^{(\pm)}$, is strongly suppressed and remains negligibly small for all $V_g$ (see Appendix B).
For intra-valley scattering, $T_\pm^{(\pm)}$, a potential barrier ($V_g>0$) is nearly opaque for $D+$ electrons [Fig. 4(b)] but highly transparent for $D-$ electrons [Fig. 4(c)]. In contrary, the valley-preference of a potential well ($V_g<0$) behaves in an opposite fashion -- transmission of electrons in $D+$ valley is preferred [Fig. 4(d)] while that of $D-$ valley is strongly suppressed [Fig. 4(e)]. This valley-selective transport is a direct consequence of the matching and mismatching of the pseudospin configurations as discussed above [Fig. 1(a)] and immediately suggests that the proposed device can be operated as a gate-tunable valley filter.  

Macroscopically, such valley-selective quantum tunneling effect manifests in the transport measurement by exhibiting valley-polarized electrical conductance. To illustrate this, we separate the contribution from $D\pm$ valleys by defining a \emph{valley-dependent electrical conductance} as $G_{\pm}(\varepsilon_F, V_g)  = G_0  \sum_{\mathcal{D}' = \pm} \int dk_y T_{\pm}^{(\mathcal{D}')}(k_y, \varepsilon_F, V_g)$ where $\varepsilon_F$ is the Fermi level of the sample, $G_0 \equiv W g_0 k_0/2\pi$, $g_0 \equiv 4e^2/h$ and $W$ is the sample width.
The $V_g$-dependence of $G_{\pm}$ is shown in Figs. 4(f) and (g). 
Apart from the expected conductance oscillations due to Fabry-P\'erot interference, it can be seen that $G_{\pm}$ dominates well separated regime of $V_g$. 
For $V_g< 0$ and $V_g>0$, electrical conduction occurs almost exclusively via $G_{+}$ and $G_{-}$, respectively, thus demonstrating a gate-tunable valley polarization of the electrical current.
Only at the vicinity of $V_g =  0$, $G_{+}$ and $G_{-}$ mixes due to the presence of both $D\pm$ transmission pathway. 
The sharp dip of $G_{\pm}$ when $V_g \approx \varepsilon_{F}$ corresponds to the case when $\varepsilon_F$ is situated at the Dirac point which has a vanishing density of states.

The valley filtering effect can be characterized by the valley polarization efficiency, 
\begin{equation}
\eta( \varepsilon_{F}, V_g) =  \frac{G_+(\varepsilon_{F}, V_g) - G_- (\varepsilon_{F}, V_g) } { G_+(\varepsilon_{F}, V_g) + G_- (\varepsilon_{F}, V_g)},
\end{equation}
which exhibits two remarkable behaviors [Fig. 4(h)]. First, high degree of valley polarization persists over a \emph{semi-infinite} $V_g$-window, indicating high robustness against noise fluctuations of $V_g$. Second, the valley polarization can be switched `off' into a stable null-polarization state by setting the gate voltage to $ - \left(\left|\Delta\right| - \varepsilon_F \right) \lesssim V_g \lesssim \left( \left|\Delta\right| + \varepsilon_F \right)$. 
Together with the two $D\pm$ polarization states, this forms (2+1) stable valley states which can be used to implement universal reversible Boolean logics as shown below.

\subsection{Valley valve and colossal valley-pseudo-magnetoresistance}

\begin{figure*}[t]
	\includegraphics[scale=.32]{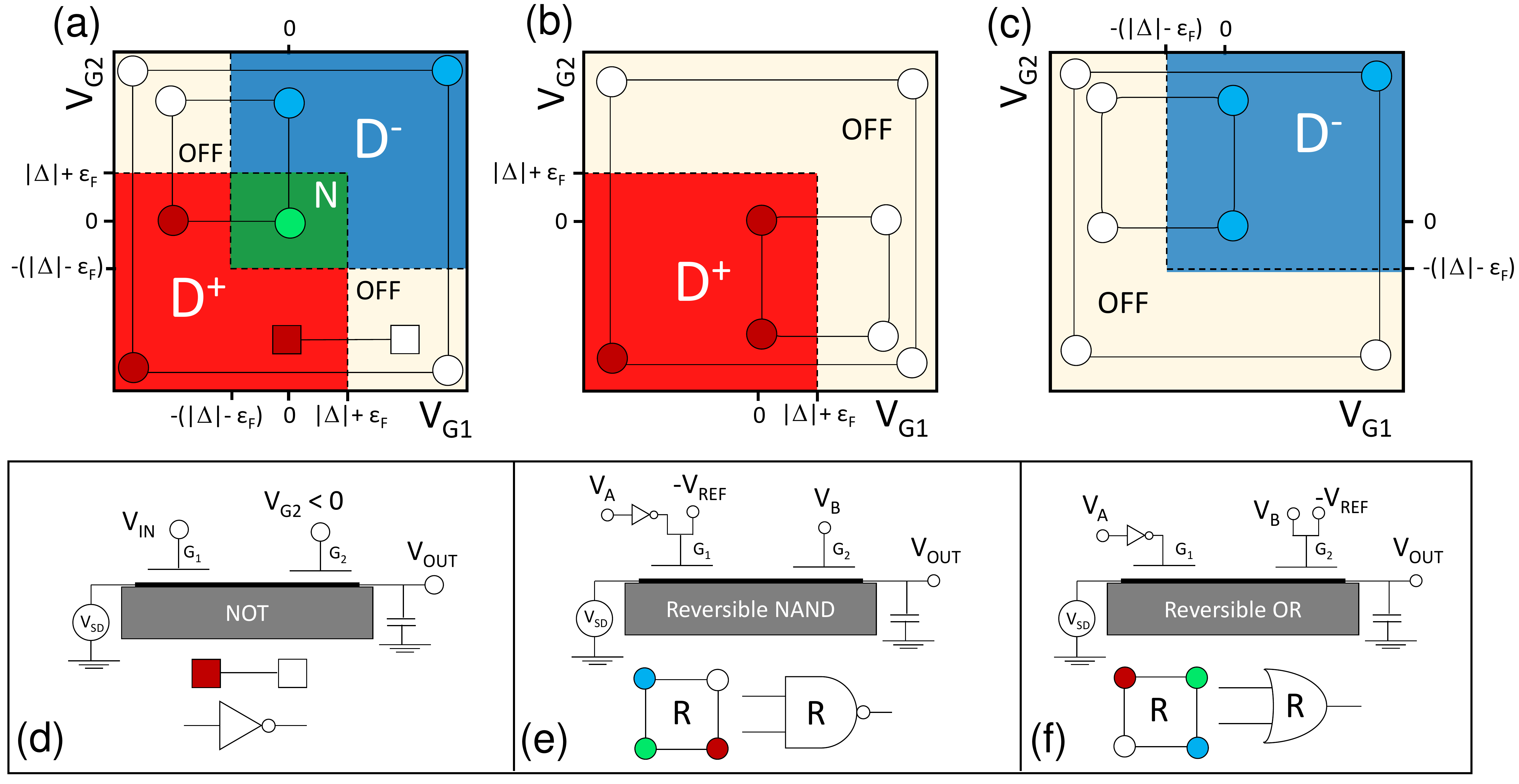}	
	\caption{Valley-transport phase diagrams (VPD) and universal reversible valleytronic logic gates. (a) VPD of the valley filter. (b) VPD for $D+$ mode and (c) $D-$ mode. One-input NOT operation is indicated by a single Boolean line in (a). The larger and the smaller loop in (a) represents Class-2A and Class-3 operations, respectively. In (b) and (c), the larger and smaller loop represents Class-1 and Class-2B operations. Physical implementation of (d) NOT, (e) reversible NAND and (f) reversible OR gates.}
\end{figure*}

We now propose a \emph{valley valve} capable of performing current on-off switching. The valley valve is composed of two gates, $V_{G1}$ and $V_{G2}$, that serve as the functional core of the valve, and a third `selector' gate $V_s$ which, in analogous to the role of `analyzer' in a optical polarizer-analyzer system, switches the valley valve into $D+$ and $D-$ mode by providing an additional stage of filtering [Fig. 5(a)]. 

The tuning of $V_{G1}$ and $V_{G2}$ creates four quadrants of parallel and anti-parallel configurations as denoted by Q1 to Q4 in Fig. 5(b).
The $D+$ and $D-$ conduction dominates Q3 and Q1, respectively, and the current can be switched off by operating the valve in Q2 and Q4. 
Figs. 5(c)-(e) shows the numerical results of $G_{\pm}(\varepsilon_F, V_{G1}, V_{G2})$ and the corresponding $\eta$. 
The conductances exhibit Fabry-P\'erot oscillations due to interference of wavefunction.
The $G_+$ and $G_-$ plateaus occurs in a \emph{semi-infinite} regimes defined by $ V_{g1,g2} < (\left|\Delta\right| + \varepsilon_F)$ and $ V_{G1,G2} > -(\left|\Delta\right| - \varepsilon_F)$, respectively. 
The intersections of $G_+$ and $G_-$ plateaus forms a central null-polarization `square' (i.e. $\eta \approx 0$) as bounded by $ - (\left|\Delta\right| - \varepsilon_F)< V_{G1,G2} < (\left|\Delta\right| + \varepsilon_F)$.
A valley valve operating in $D+$ mode can be obtained by switching $V_{G2}$ between Q2$\leftrightarrows$Q3 while fixing $V_{G1}$ at a negative value, or by switching $V_{G1}$ between Q3$\leftrightarrows$Q4 while fixing $V_{G2}$ at a negative value. 
Similarly, a $D-$ mode valley valve can be obtained via the switching of $V_{G1}$ ($V_{G2}$) with Q1$\leftrightarrows$Q2 while fixing $V_{G2}$ at a positive value, or equivalently via the switching of $V_{G2}$ with Q1$\leftrightarrows$Q4 while fixing $V_{G1}$ at a positive value.
In Figs 5(d) and (e), the $D-$ and $D+$ conduction blocks is selectively suppressed by setting $V_s = -0.2$ and $V_s = 0.2$, respectively. This switches the valley valve into an exclusively $D+$ or $D-$ modes.

The valley valve can be characterized by a \emph{tunneling valley-pseudo-magnetoresistance} (VPMR) ratio, analogous to the tunneling magnetoresistance (TMR) in magnetic tunnel junction, which is defined as
\begin{equation}
\text{VPMR} = \frac{\bar{G}_{\text{total}}^{P} - \bar{G}_{\text{total}}^{AP} }{ \bar{G}_{\text{total}}^{P} },
\end{equation}
where $\bar{G}_{\text{total}}^{P}$ and $\bar{G}_{\text{total}}^{AP}$ represents the total conductance averaged over a range of $V_{G1,2}$ in parallel and anti-parallel configuration. The VPMR for $D-$ conductance block on/off switching, i.e. Q1$\leftrightarrows$Q2, that of the $D+$ conductance block, i.e. Q3$\leftrightarrows$Q2, is shown in Fig. 5(f) with $V_s = 0$.
For $\varepsilon_{F} \to \left|\Delta \right|$, the VPMR of Q3$\leftrightarrows$Q2 gradually stabilizes at $\sim400\%$ while the VPMR of Q1$\leftrightarrows$Q2 is severely degraded due to the disappearance of $D-$ conductance block when the two Dirac cones merge.
Remarkably, the $\text{VPMR}$ exhibits a colossal value of well over $10,000\%$ at small $\varepsilon_F$. This value greatly exceeds the pseudo-magnetoresistance (PMR) of $\sim100\%$ in graphene-based pseudospin valve \cite{san-jose} and $\text{TMR}\sim1,000\%$ in traditional magnetic tunnel junction \cite{TMR}, and is on par with state-of-art tunneling electroresistance (TER) of up to $\text{TER}\sim10,000\%$ in ferroelectric tunnel junctions \cite{TER}. This colossal VPMR originates from the pseudospin-assisted tunneling described above, which effectively quenches the conduction current when there is a mismatch of pseudospin.

\subsection{Universal reversible valleytronic logic gate}

We now show that the existence of $D\pm$ blocks and the central null-polarization square in the conductance spectrum in Figs. 5(c)-(e) allows the valley valve device to be operated as a two-input Boolean combinational logic gate. 
We first translate this conductance spectrum into a simplified \emph{valley-transport phase diagram} (VPD) [Figs. 6(a)-(c)]. 
In Fig. 6(a), selector voltage is set to $V_s =0$, i.e. both $D\pm$ channels are opened. 
In this case, both the $D+$ (red) and the $D-$ (blue) blocks of conductance plateau are presence and their intersection forms a central green block of null-polarization. 
The VPDs in Figs. 6(b) and (c), corresponds the case of $D+$ and $D-$ modes by, respectively, setting $V_s =-0.2$ and $V_s = 0.2$. 

To demonstrate the logical operation of the proposed valleytronic gate, we employ a graphical \emph{Boolean loop} analysis based on the Karnaugh's map approach \cite{karnaugh}. Such method provides a simplified tool as it directly maps abstract Boolean logical operations onto the conductance spectrum of a physical system. For example, NOT can be represented by a one-dimensional \emph{Boolean line} where the node at its two edges represents the two input state of `0' and `1', whereas the output state is denoted by empty and filled node for `0' and `1' output state, respectively, i.e.
\begin{equation}
\begin{tikzpicture}[baseline=-0.65ex,scale=0.5]
\node[draw, fill = red,scale=0.5] (LM) at ( -.5,   0) {};
\node[draw,scale=0.5] (RM)   at (.5,0) {};
\draw
(LM) node [xshift = -0.2cm, yshift=0.2cm] { \scriptsize{(0)} } -- (RM) node [xshift = 0.2cm, yshift=0.2cm]{ \scriptsize{(1)} }  node [midway,yshift=-0.4cm] { \scriptsize{NOT} };
\end{tikzpicture},
\end{equation}
which can be located at the lower quadrant of VPDs in Fig. 6(a). The physical implementation of NOT can be determined as followed: the Boolean line is horizontally aligned along a constant level of negative $V_{G2}$ and the switching of $V_{G1}$ from zero to positive value changes the conductance from $D+$ block to `OFF' state. Correspondingly, by fixing gate $G_2$ at a negative reference voltage and feeding the input signal into gate $G_1$, NOT operation is obtained [see Fig. 6(d)].

We now employ this graphical method to extract the permissible two-input Boolean operations, represented by Boolean loops, from the VPDs. We use the designation of `\emph{Class-X}' to catalog all 16-types of Boolean logical operations where $X = 0,1,2,3,4$ denotes the number of filled nodes in the Boolean loop. We first define \emph{Class-0} and \emph{Class-4} logic where the four nodes at the vertices are either all empty or all filled. This corresponds to the trivial operations of `always-ON' and `always-OFF',
%
%
\begin{equation}
\begin{tikzpicture}[baseline=-0.65ex,scale=0.5]
\node[draw, circle, scale=0.5] (LB) at ( -.5,   -.5) {};
\node[draw, circle, scale=0.5] (LT)   at (-.5,.5) {};
\node[draw, circle, scale=0.5] (RB)  at ( .5, -.5) {};
\node[draw, circle, scale=0.5] (RT)    at ( .5,   .5) {};
\draw
(LB) node [xshift=-0.5cm] { \scriptsize{(0,0)} } -- (LT)node [xshift=-0.5cm] { \scriptsize{(0,1)} }  -- (RT) node [xshift=0.5cm] { \scriptsize{(1,1)} }  -- (RB)  node [xshift=0.5cm] { \scriptsize{(1,0)} } -- (LB) node [midway,yshift=-0.4cm] { \scriptsize{Always-OFF} };
\end{tikzpicture}
\text{  ,  }
\begin{tikzpicture}[baseline=-0.65ex,scale=0.5]
\node[draw, circle, fill = red, scale=0.5] (LB) at ( -.5,   -.5) {} ;
\node[draw, circle, fill = red, scale=0.5] (LT)   at (-.5,.5) {};
\node[draw, circle, fill = red, scale=0.5] (RB)  at ( .5, -.5) {};
\node[draw, circle, fill = red, scale=0.5] (RT)    at ( .5,   .5) {};
\draw
(LB) -- (LT) -- (RT) -- (RB) -- (LB) node [midway,yshift=-0.4cm] { \scriptsize{Always-ON} };
\end{tikzpicture}
, 
\end{equation}
which can be implemented by drawing a Boolean loop lying completely outside and inside the $D_{\pm}$ conductance blocks, respectively. Note that the input address, $(A,B)$, is explicitly marked in the `always-OFF' Boolean loop and is omitted in the following discussion for simplicity.

\emph{Class-1} logics, in which the Boolean loop contains only one filled-node, can be implemented via the large Boolean loops shown in the VPDs of $D+$ [Fig. 6(b)] and $D-$ [Fig. 6(c)] mode:
%
%
\begin{subequations}
	\begin{equation}
	\begin{tikzpicture}[baseline=-0.65ex,scale=0.5]
	\node[draw, circle, fill = red, scale=0.5] (LB) at ( -.5,   -.5) {};
	\node[draw, circle, scale=0.5] (LT)   at (-.5,.5) {};
	\node[draw, circle, scale=0.5] (RB)  at ( .5, -.5) {};
	\node[draw, circle,scale=0.5] (RT)    at ( .5,   .5) {};
	\draw
	(LB) -- (LT) -- (RT) -- (RB) -- (LB) node [midway,yshift=-0.4cm] { \scriptsize{NOR} };
	\end{tikzpicture}
	\text{  $\circlearrowright$ }
	\begin{tikzpicture}[baseline=-0.65ex,scale=0.5]
	\node[draw, circle, scale=0.5] (LB) at ( -.5,   -.5) {};
	\node[draw, circle, fill = red, scale=0.5] (LT)   at (-.5,.5) {};
	\node[draw, circle, scale=0.5] (RB)  at ( .5, -.5) {};
	\node[draw, circle,scale=0.5] (RT)    at ( .5,   .5) {};
	\draw
	(LB) -- (LT) -- (RT) -- (RB) -- (LB) node [midway,yshift=-0.4cm] { \scriptsize{IMPLY} };
	\end{tikzpicture}
	\text{ $\circlearrowright$ }
	\begin{tikzpicture}[baseline=-0.65ex,scale=0.5]
	\node[draw, circle, scale=0.5] (LB) at ( -.5,   -.5) {};
	\node[draw, circle, scale=0.5] (LT)   at (-.5, .5) {};
	\node[draw, circle, scale=0.5] (RB)  at ( .5, -.5) {};
	\node[draw, circle, fill=red, scale=0.5] (RT)    at ( .5,   .5) {};
	\draw
	(LB) -- (LT) -- (RT) -- (RB) -- (LB) node [midway,yshift=-0.4cm] { \scriptsize{AND} };
	\end{tikzpicture}
	\text{ $\circlearrowright$ }
	\begin{tikzpicture}[baseline=-0.65ex,scale=0.5]
	\node[draw, circle, scale=0.5] (LB) at ( -.5,   -.5) {};
	\node[draw, circle, scale=0.5] (LT)   at (-.5,.5) {};
	\node[draw, circle, fill = red,scale=0.5] (RB)  at ( .5, -.5) {};
	\node[draw, circle, scale=0.5] (RT)    at ( .5,   .5) {};
	\draw
	(LB) -- (LT) -- (RT) -- (RB) -- (LB) node [midway,yshift=-0.4cm] { \scriptsize{C-IMPLY} };
	\end{tikzpicture},
	\end{equation}
	\begin{equation}
	\begin{tikzpicture}[baseline=-0.65ex,scale=0.5]
	\node[draw, circle, scale=0.5] (LB) at ( -.5,   -.5) {};
	\node[draw, circle, scale=0.5] (LT)   at (-.5,.5) {};
	\node[draw, circle, scale=0.5] (RB)  at ( .5, -.5) {};
	\node[draw, circle, fill=blue, scale=0.5] (RT)    at ( .5,   .5) {};
	\draw
	(LB) -- (LT) -- (RT) -- (RB) -- (LB) node [midway,yshift=-0.4cm] { \scriptsize{AND} };
	\end{tikzpicture}
	\text{  $\circlearrowright$ }
	\begin{tikzpicture}[baseline=-0.65ex,scale=0.5]
	\node[draw, circle, scale=0.5] (LB) at ( -.5,   -.5) {};
	\node[draw, circle, scale=0.5] (LT)   at (-.5,.5) {};
	\node[draw, circle, fill = blue, scale=0.5] (RB)  at ( .5, -.5) {};
	\node[draw, circle, scale=0.5] (RT)    at ( .5,   .5) {};
	\draw
	(LB) -- (LT) -- (RT) -- (RB) -- (LB) node [midway,yshift=-0.4cm] { \scriptsize{IMPLY} };
	\end{tikzpicture}
	\text{ $\circlearrowright$ }
	\begin{tikzpicture}[baseline=-0.65ex,scale=0.5]
	\node[draw, circle, fill=blue, scale=0.5] (LB) at ( -.5,   -.5) {};
	\node[draw, circle, scale=0.5] (LT)   at (-.5, .5) {};
	\node[draw, circle, scale=0.5] (RB)  at ( .5, -.5) {};
	\node[draw, circle, scale=0.5] (RT)    at ( .5,   .5) {};
	\draw
	(LB) -- (LT) -- (RT) -- (RB) -- (LB) node [midway,yshift=-0.4cm] { \scriptsize{NOR} };
	\end{tikzpicture}
	\text{ $\circlearrowright$ }
	\begin{tikzpicture}[baseline=-0.65ex,scale=0.5]
	\node[draw, circle, scale=0.5] (LB) at ( -.5,   -.5) {};
	\node[draw, circle, fill=blue, scale=0.5] (LT)   at (-.5,.5) {};
	\node[draw, circle, scale=0.5] (RB)  at ( .5, -.5) {};
	\node[draw, circle, scale=0.5] (RT)    at ( .5,   .5) {};
	\draw
	(LB) -- (LT) -- (RT) -- (RB) -- (LB) node [midway,yshift=-0.4cm] { \scriptsize{C-IMPLY} };
	\end{tikzpicture},
	\end{equation}
\end{subequations}
Importantly, the universal NOR gate falls into this class and can be implemented in $D+$ mode. 
Moreover, the exotic implication-type operations, such as the negations of implication (N-IMPLY) and of converse-implication (NC-IMPLY) can also be obtained via \emph{circular permutations} of the Boolean loop in $V_{G1}$-$V_{G2}$-space [denoted by `$\circlearrowright$' in Eq. (6)] and can be physically implemented by properly inverting and/or combining reference voltages with the voltage signals [see Figs. 6(d)-(f) for examples].

The smaller Boolean loops shown in Figs. 6(b) and (c) represent Class-2A logics of two consecutive filled nodes:
%
%
\begin{subequations}
	\begin{equation}
	\begin{tikzpicture}[baseline=-0.65ex,scale=0.5]
	\node[draw, circle, fill = red, scale=0.5] (LB) at ( -.5,   -.5) {};
	\node[draw, circle, fill = red, scale=0.5] (LT)   at (-.5,.5) {};
	\node[draw, circle, scale=0.5] (RB)  at ( .5, -.5) {};
	\node[draw, circle, scale=0.5] (RT)    at ( .5,   .5) {};
	\draw
	(LB) -- (LT) -- (RT) -- (RB) -- (LB) node [midway,yshift=-0.4cm] {\scriptsize{NOT-A} };
	\end{tikzpicture}
	\text{  $\circlearrowright$ }
	\begin{tikzpicture}[baseline=-0.65ex,scale=0.5]
	\node[draw, circle, scale=0.5] (LB) at ( -.5,   -.5) {};
	\node[draw, circle, fill=red, scale=0.5] (LT)   at (-.5,.5) {};
	\node[draw, circle, scale=0.5] (RB)  at ( .5, -.5) {};
	\node[draw, circle, fill=red, scale=0.5] (RT)    at ( .5,   .5) {};
	\draw
	(LB) -- (LT) -- (RT) -- (RB) -- (LB) node [midway,yshift=-0.4cm] { \scriptsize{B} };
	\end{tikzpicture}
	\text{ $\circlearrowright$ }
	\begin{tikzpicture}[baseline=-0.65ex,scale=0.5]
	\node[draw, circle, scale=0.5] (LB) at ( -.5,   -.5) {};
	\node[draw, circle, scale=0.5] (LT)   at (-.5, .5) {};
	\node[draw, circle, fill=red, scale=0.5] (RB)  at ( .5, -.5) {};
	\node[draw, circle, fill=red, scale=0.5] (RT)    at ( .5,   .5) {};
	\draw
	(LB) -- (LT) -- (RT) -- (RB) -- (LB) node [midway,yshift=-0.4cm] { \scriptsize{A} };
	\end{tikzpicture}
	\text{ $\circlearrowright$ }
	\begin{tikzpicture}[baseline=-0.65ex,scale=0.5]
	\node[draw, circle, fill=red, scale=0.5] (LB) at ( -.5,   -.5) {};
	\node[draw, circle, scale=0.5] (LT)   at (-.5, .5) {};
	\node[draw, circle, fill=red, scale=0.5] (RB)  at ( .5, -.5) {};
	\node[draw, circle, scale=0.5] (RT)    at ( .5,   .5) {};
	\draw
	(LB) -- (LT) -- (RT) -- (RB) -- (LB) node [midway,yshift=-0.4cm] { \scriptsize{NOT-B} };
	\end{tikzpicture}
	\end{equation}
	\begin{equation}
	\begin{tikzpicture}[baseline=-0.65ex,scale=0.5]
	\node[draw, circle, scale=0.5] (LB) at ( -.5,   -.5) {};
	\node[draw, circle, scale=0.5] (LT)   at (-.5, .5) {};
	\node[draw, circle, fill=blue, scale=0.5] (RB)  at ( .5, -.5) {};
	\node[draw, circle, fill=blue, scale=0.5] (RT)    at ( .5,   .5) {};
	\draw
	(LB) -- (LT) -- (RT) -- (RB) -- (LB) node [midway,yshift=-0.4cm] { \scriptsize{A} };
	\end{tikzpicture}
	\text{ $\circlearrowright$ }
	\begin{tikzpicture}[baseline=-0.65ex,scale=0.5]
	\node[draw, circle, fill=blue, scale=0.5] (LB) at ( -.5,   -.5) {};
	\node[draw, circle, scale=0.5] (LT)   at (-.5, .5) {};
	\node[draw, circle, fill=blue, scale=0.5] (RB)  at ( .5, -.5) {};
	\node[draw, circle, scale=0.5] (RT)    at ( .5,   .5) {};
	\draw
	(LB) -- (LT) -- (RT) -- (RB) -- (LB) node [midway,yshift=-0.4cm] { \scriptsize{NOT-B} };
	\end{tikzpicture}
	\text{ $\circlearrowright$ }
	\begin{tikzpicture}[baseline=-0.65ex,scale=0.5]
	\node[draw, circle, fill = blue, scale=0.5] (LB) at ( -.5,   -.5) {};
	\node[draw, circle, fill = blue, scale=0.5] (LT)   at (-.5,.5) {};
	\node[draw, circle, scale=0.5] (RB)  at ( .5, -.5) {};
	\node[draw, circle, scale=0.5] (RT)    at ( .5,   .5) {};
	\draw
	(LB) -- (LT) -- (RT) -- (RB) -- (LB) node [midway,yshift=-0.4cm] { \scriptsize{NOT-A} };
	\end{tikzpicture}
	\text{  $\circlearrowright$ }
	\begin{tikzpicture}[baseline=-0.65ex,scale=0.5]
	\node[draw, circle, scale=0.5] (LB) at ( -.5,   -.5) {};
	\node[draw, circle, fill=blue, scale=0.5] (LT)   at (-.5,.5) {};
	\node[draw, circle, scale=0.5] (RB)  at ( .5, -.5) {};
	\node[draw, circle, fill=blue, scale=0.5] (RT)    at ( .5,   .5) {};
	\draw
	(LB) -- (LT) -- (RT) -- (RB) -- (LB) node [midway,yshift=-0.4cm] { \scriptsize{B} };
	\end{tikzpicture},
	\end{equation}
\end{subequations}
which are rather trivial logical operations.
\emph{Class-2B} logics, in which the two filled nodes are separated, provides more useful operations of XOR and XNOR. This class is obtainable from the larger Boolean loop shown in Fig. 6(a):
%
\begin{equation}
\begin{tikzpicture}[baseline=-0.65ex,scale=0.5]
\node[draw, circle, fill = red, scale=0.5] (LB) at ( -.5,   -.5) {};
\node[draw, circle, scale=0.5] (LT)   at (-.5,.5) {};
\node[draw, circle, scale=0.5] (RB)  at ( .5, -.5) {};
\node[draw, circle, fill = blue, scale=0.5] (RT)    at ( .5,   .5) {};
\draw
(LB) -- (LT) -- (RT) -- (RB) -- (LB) node [midway,yshift=-0.4cm] { \scriptsize{XNOR} };
\end{tikzpicture}
\text{  $\circlearrowright$ }
\begin{tikzpicture}[baseline=-0.65ex,scale=0.5]
\node[draw, circle, scale=0.5] (LB) at ( -.5,   -.5) {};
\node[draw, circle, fill=red, scale=0.5] (LT)   at (-.5,.5) {};
\node[draw, circle, fill = blue, scale=0.5] (RB)  at ( .5, -.5) {};
\node[draw, circle, scale=0.5] (RT)    at ( .5,   .5) {};
\draw
(LB) -- (LT) -- (RT) -- (RB) -- (LB) node [midway,yshift=-0.4cm] { \scriptsize{XOR} };
\end{tikzpicture}
\end{equation}
Interestingly, the two logical `ON' output states are contains opposite valley polarizations as denoted by blue and red nodes.
Such \emph{valley-color labeling} manifests in a more remarkable way for \emph{Class-3} logics where three nodes are filled. This class is represented in the smaller Boolean loop in Fig. 6(a), i.e.
%
%
\begin{equation}
\begin{tikzpicture}[baseline=-0.65ex,scale=0.5]
\node[draw, circle,fill = blue, scale=0.5] (LB) at ( -.5,   -.5) {};
\node[draw, circle, scale=0.5] (LT)   at (-.5,.5) {};
\node[draw, circle, fill = ForestGreen, scale=0.5] (RB)  at ( .5, -.5) {};
\node[draw, circle, fill = red, scale=0.5] (RT)    at ( .5,   .5) {};
\draw
(LB) -- (LT) -- (RT) -- (RB) -- (LB) node [midway,yshift=-0.4cm] { \scriptsize{C-IMPLY} } node [midway,yshift=0.25cm] { \footnotesize{R} };
\end{tikzpicture}
\text{  $\circlearrowright$ }
\begin{tikzpicture}[baseline=-0.65ex,scale=0.5]
\node[draw, circle, fill = ForestGreen, scale=0.5] (LB) at ( -.5,   -.5) {};
\node[draw, circle, fill = blue, scale=0.5] (LT)   at (-.5,.5) {};
\node[draw, circle, fill = red, scale=0.5] (RB)  at ( .5, -.5) {};
\node[draw, circle, scale=0.5] (RT)    at ( .5,   .5) {};
\draw
(LB) -- (LT) -- (RT) -- (RB) -- (LB) node [midway,yshift=-0.4cm] { \scriptsize{NAND} } node [midway,yshift=0.25cm] { \footnotesize{R} };
\end{tikzpicture}
\text{ $\circlearrowright$ }
\begin{tikzpicture}[baseline=-0.65ex,scale=0.5]
\node[draw, circle, fill = red, scale=0.5] (LB) at ( -.5,   -.5) {};
\node[draw, circle, fill = ForestGreen, scale=0.5] (LT)   at (-.5,.5) {};
\node[draw, circle, scale=0.5] (RB)  at ( .5, -.5) {};
\node[draw, circle, fill = blue, scale=0.5] (RT)    at ( .5,   .5) {};
\draw
(LB) -- (LT) -- (RT) -- (RB) -- (LB) node [midway,yshift=-0.4cm] { \scriptsize{IMPLY} } node [midway,yshift=0.25cm] { \footnotesize{R} };
\end{tikzpicture}
\text{ $\circlearrowright$ }
\begin{tikzpicture}[baseline=-0.65ex,scale=0.5]
\node[draw, circle, scale=0.5] (LB) at ( -.5,   -.5) {};
\node[draw, circle, fill = red, scale=0.5] (LT)   at (-.5,.5) {};
\node[draw, circle, fill = ForestGreen, scale=0.5] (RB)  at ( .5, -.5) {};
\node[draw, circle, fill = blue, scale=0.5] (RT)    at ( .5,   .5) {};
\draw
(LB) -- (LT) -- (RT) -- (RB) -- (LB) node [midway,yshift=-0.4cm] { \scriptsize{OR} } node [midway,yshift=0.25cm] { \footnotesize{R} };
\end{tikzpicture}.
\end{equation}
Eq. (4)-(9) demonstrate that the valleytronic logic gates proposed here are capable of hosting all 16 types of two-input Boolean combinational logics.
The Class 3 logics in Eq. (9) includes exotic implication (IMPLY), converse-implication (C-IMPLY) operations and the conventional OR and NAND.
NAND can be obtained via one circular permutation, i.e. by inverting one input and subsequently referencing it with a negative reference voltage as shown in Fig. 6(e). 
For OR gate, three circular permutations are required, which correspond to inverting the first input and referencing the second input with a negative voltage [Fig. 6(f)].

Class-3 logics are logically-reversible as each of the three filled-nodes is unambiguously labeled by (2+1) valley polarization states. 
By reading out the valley polarization, any output can be unambiguously reverted back to their corresponding initial input. 
More remarkably, this reversible class of logics includes the all-important universal NAND gate. This reveals a remarkable potential of the valleytronic logical gate proposed here as a building block of classical reversible computer.

\begin{figure}[t]
	\includegraphics[scale=.305]{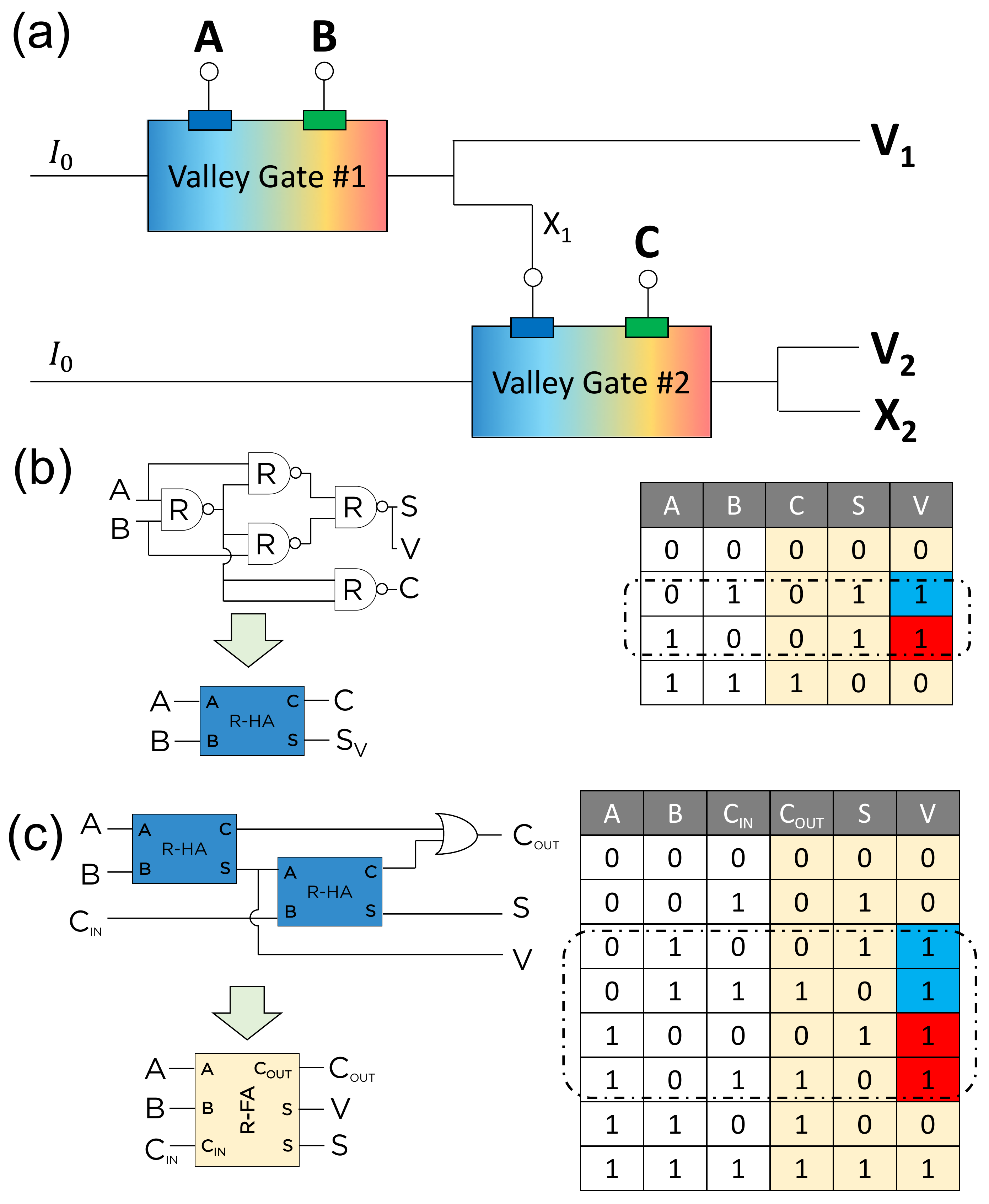}	
	\caption{Examples of valleytronic-based reversible Boolean circuit. (a) General scheme of a valleytronic-based reversible circuit. Three inputs, (\textbf{A}, \textbf{B}, \textbf{C}), are operated by two gates to yield a final current output, $X_2$. The input states can be unambiguously recovered via the two valley polarizations, $V_1$ and $V_2$. (b) Reversible half-adder and (c) reversible full-adder with their truth table. Ambiguous operations are emphasized by dashed boxes. In these devices, the logical-reversibility is established via the valley colors, $V$.}
\end{figure}

A general scheme of valleytronic-based reversible Boolean circuits is illustrated in Fig. 7(a). The three inputs, (\textbf{A}, \textbf{B}, \textbf{C}), are computed reversibly into a final output $X_2$ in this example. A direct current, $I_0$, is fed into each of the two-input valley gates and is modulated by the two inputs to yield an output current of a particular state of valley polarization. For `valley gate \#1', \textbf{A} and \textbf{B} yield an intermediate output current, $X_1$, of valley polarization $V_1$. $X_1$ is subsequently cascaded into `valley gate \#2', and its combination with the third input \textbf{C} produces the final output current, $X_2$, of valley polarization, $V_2$. For a given $X_2$, the initial input states can be unambiguously recovered from the $V_1$ and $V_2$. 
Based on this general scheme, \emph{reversible half-adder} (R-HA) can be implemented as shown in Fig. 7(b). From its truth table, it can be seen that the output state of `1' that corresponds to ambiguous input states of $(A,B) = \{(1,0), (0,1)\}$ can be uniquely distinguishable from the valley polarization of the output [represented by $V$ in the truth table of Fig. 7(b)]. Two units of R-HA can be cascaded into a \emph{reversible full-adder} (R-FA) [Fig. 7(c)]. In this case, the input states of $(C_{\text{IN}},A, B) = \{(1,0,0), (0,1,0), (0,0,1)\}$ can be unambiguously distinguished via the valley polarization, $V$, of the intermediate $C$ output.

It should be emphasized that the valleytronic-based reversible circuit is fundamentally different from that of the traditional approach.
Traditional reversible gates, such as Fredkin and Tofolli, achieve logical-reversibility via multiple supplementary bits \cite{toffoli, fredkin} which inevitably introduce \emph{ancilla inputs} and \emph{garbage outputs} -- bits unrelated to computation results and are only required for logical-reversibility -- into the circuits.
In contrast, the reversible valleytronic logic gates proposed here harnesses the built-in valley degree of freedom for information storage.
Extraction of information encoded in the valley is performed via an \emph{on-demand} fashion, i.e. valley polarization is read-only only when it is absolutely needed for logical reversibility. 
More importantly, the valleytronic reversible gate retains the conventional two-input format without involving any ancilla inputs.
Thus, the combination of (i) reduced garbage outputs; (ii) complete absence of ancilla inputs; and (iii) the retaining of conventional two-input format suggests that the valleytronic-approach for reversible computation can potentially be more advantageous than the traditional approach. 
We further note that although logical-reversibility can break the Landauer's limit, it does not warrant the full elimination of energy dissipation. 
Energy dissipation in a reversible computer shall remain finite due to the inevitable \emph{physical-irreversibility} of electronic devices and circuits \cite{lloyd, markov}.
Finally, we remark that it remains a technological challenge to fabricate the proposed valleytronic devices at current stage. The search for an ideal merging Dirac cone condensed matter system shall form an on-going task before the full potential of valleytronic-based logic gate architecture proposed here can be tapped. 

\section{Conclusion}

In summary, we study the pseudospin-assisted valley-contrasting quantum tunneling in 2MDS and show that such effect can be harnessed to create valley filter, valve and logic gate.
These valleytronic devices exhibit multiple unusual characteristics including: (i) all-electric controllable; (ii) stable valley polarization that persists over semi-infinite gate voltage windows; (iii) colossal VPMR effect of well over 10,000\%; (iv) flexibility to be permuted into any two-input Boolean gates; and (v) capable of performing reversible Boolean classical computation with reduced garbage and total absence of ancilla bits. 
The union of valley degree of freedom and digital computing offers an exciting solid state platform for valleytronic-based information processing and for reversible computing which is ultimately required to lower hardware power consumption beyond the bound of Landauer's limit \cite{saeedi}.
As logical-reversibility is a prerequisite for quantum gate \cite{rev}, we anticipate the universal reversible valleytronic logic gates proposed here to play a role in quantum \cite{QC} and quantum-classical hybrid computers \cite{hybrid}.

\begin{acknowledgments}

We thank K. J. A. Ooi, M. Zubair, and S.-H. Tan for helpful discussion. This work is funded by A*STAR-IRG (A1783c0011) and Singapore Ministry of Education MOE-Tier-2 Grant (T2MOE1401).

\end{acknowledgments}

\begin{appendix}
	
\section{Chirality of the merging Dirac cones}

In this section, we show that the two Dirac split cones predicted by the universal Hamiltonian [Eq. (1) of main text] for $\Delta < 0$ contains opposite chirality. We first expand the Hamiltonian around the two Dirac points, i.e. $\mathbf{k} \to \tilde{\mathbf{k}}$ where $\tilde{ \mathbf{k} } = (\mathcal{D} \sqrt{\left| \Delta \right|} +\delta k_x,  \delta k_y)$, $\delta \mathbf{k} \equiv (\delta k_x, \delta k_y)$ is a small shift of wavevector and $\mathcal{D} = \pm 1$ denotes the two Dirac cones. By keeping terms first order in $(\delta k_x, \delta k_y)$, the dimensionless Hamiltonian becomes can be decoupled for each of the Dirac cone, i.e.
\begin{equation}
\hat{\mathcal{H}}^{(\mathcal{D})}_{\delta \mathbf{k}} = 
\begin{pmatrix}
0 & \mathcal{D} \sqrt{\left| \Delta \right|} \delta k_x - i\delta k_y \\
\mathcal{D} \sqrt{\left| \Delta \right|} \delta k_x + i\delta k_y & 0
\end{pmatrix}
\end{equation}
which coincides with the gapless Dirac Hamiltonian of graphene except that the `Fermi velocity' is anisotropic between $\delta k_x$ and $\delta k_y$ directions. The chirality operators can be defined as $\hat{\chi}_{\mathcal{D}} = \left(\vec{\mathcal{K}}_{\mathcal{D}}/ \left| \vec{\mathcal{K}}_{\mathcal{D}} \right|\right) \cdot \hat{\sigma}$ where $ \vec{\mathcal{K}}_{\mathcal{D}} \equiv (\mathcal{D}\sqrt{\left|\Delta\right|}\delta k_x, \delta k_y)$ is an `anisotropic wavevector'. By defining a phase factor as $\phi_{\delta \mathbf{k}}  \equiv \tan^{-1} \left( \delta k_y / \sqrt{\left| \Delta \right|} \delta k_x \right)$, the eigen-energy and eigenstate can be solved, respectively, as $\varepsilon_{\delta \mathbf{k} ,s} = s \left| \vec{\mathcal{K}}_{\mathcal{D}} \right|$ and $\ket{ \psi_{\delta \mathbf{k}, \mathcal{D}} } = 
\left(1, se^{i\mathcal{D} \phi_{\delta \mathbf{k}}}\right)^{\mathcal{T}}e^{i \delta \mathbf{k} \cdot \mathbf{r}}$ where $\mathbf{r} \equiv (x,y)$ and $s = \pm 1$ is the band index. $\hat{\chi}_\mathcal{D}$ commutes with $\hat{\mathcal{H}}_{\delta \mathbf{k}}^{(\mathcal{D})}$ and follows the eigenvalue equation: $\hat{\chi}_{\mathcal{D}} \ket{ \psi_{\delta \mathbf{k}, \mathcal{D}} } = \chi_\mathcal{D}  \ket{ \psi_{\delta \mathbf{k}, \mathcal{D}} }$, where the chirality eigenvalue is $\chi_\mathcal{D} = s \mathcal{D}$. This demonstrate that $\chi_\mathcal{D}$ has opposite sign between $\mathcal{D} = \pm 1$ valleys.

\begin{figure} [t]
	\includegraphics[scale= .65]{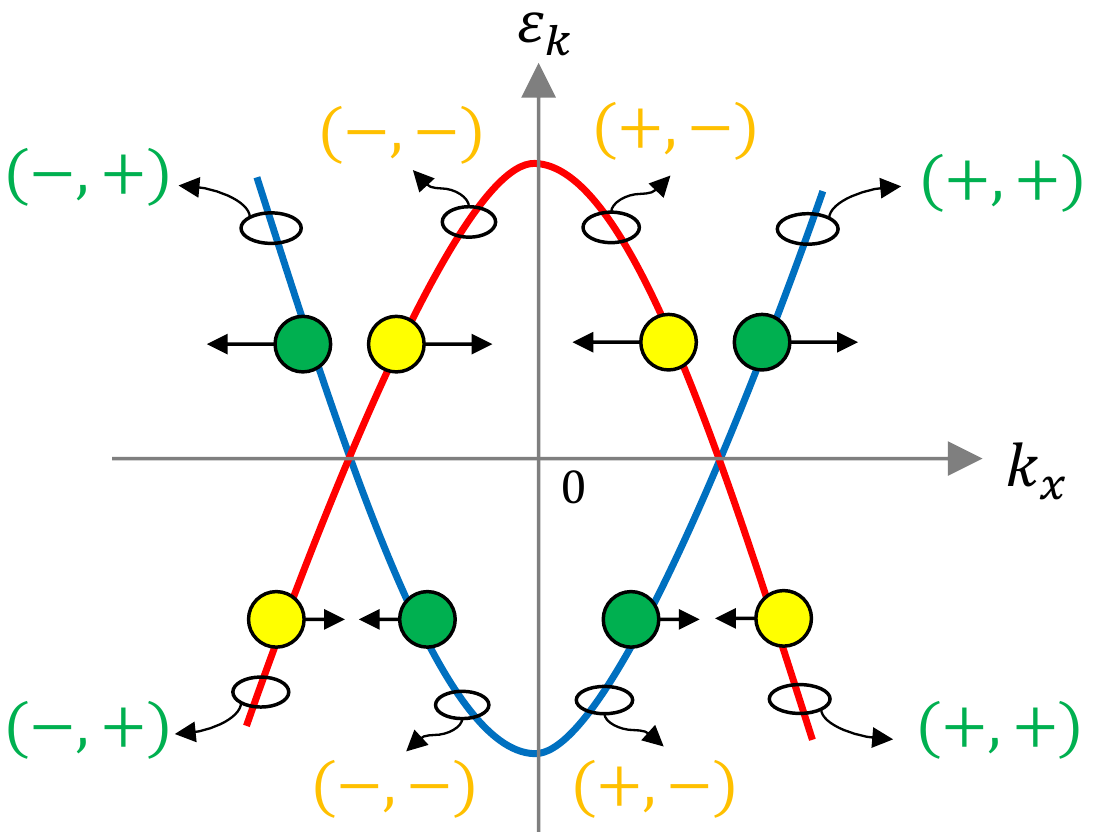}
	\caption{Energy dispersion of merging Dirac cone system at $k_y = 0$. The $(\lambda, \eta)$ index of various branches are marked. The arrows denote the direction of the group velocity, $v_x^{(\lambda\eta)}$. Electron-like ($v_x^{(\lambda\eta)}$ parallel with $k_x$) and hole-like ($v_x^{(\lambda\eta)}$ anti-parallel with $k_x$) quasiparticles are denoted by green and yellow circles, respectively. }
\end{figure}	
	
\section{Derivation of tunneling conductance of chiral valley filter}

We consider quantum tunneling across a 1D square potential along the $x$-direction:
\begin{equation}
U(x) = V_g \left[ \Theta\left(x\right) - \Theta\left(x-d\right)
\right]
\end{equation}
where $V_g \equiv e\mathcal{V}_g / \varepsilon_0$ is a dimensionless potential barrier/well height determined by the gate voltage $\mathcal{V}_g$ and $d = k_0 d_0$ is the dimensionless width parameter that corresponds to the barrier width, $d_0$. By replacing $k_x \to -i \partial/\partial_x$, the Schr\"odinger can be written explicitly as
\begin{equation}
\begin{pmatrix}
0 & -\frac{\partial^2}{\partial x^2} + \Delta - ik_y \\
-\frac{\partial^2}{\partial x^2} + \Delta + ik_y
\end{pmatrix}
\psi(x) = \left(\varepsilon_k - U(x) \right) \psi(x)
\end{equation}
which can be decoupled as
\begin{equation}
\left(\frac{\partial^4}{\partial x^4} - 2\Delta^2 \frac{\partial^2}{\partial x^2}  \right) \phi_{A,B} = \left(\left(\varepsilon_k - U(x)\right)^2-k_y^2 - \Delta^2 \right) \phi_{A,B}
\end{equation}
where the $\psi(x) = (\phi_A, \phi_B)^\mathcal{T}$ is the pseudospinor wavevefunction. The solutions of the first pseudospinor-component with $U(x) = 0$ can be solved as $\phi_A^{(\lambda\eta)} = \exp\left(ik_x^{(\lambda\eta)} x\right)$ where $\lambda = \pm 1$, $\eta = \pm 1$,
\begin{equation}
k_x^{(\lambda\eta)} = \lambda \sqrt{ \eta \left(\varepsilon_k^2 - k_y^2\right)^{1/2} - \Delta }, 
\end{equation}
and the energy eigenvalue is given as
\begin{equation}
\varepsilon_{\mathbf{k}} = \pm \sqrt{ \left(k_x^{(\lambda\eta)2} + \Delta \right)^2 + k_y^2}
\end{equation}
For $\Delta < 0$ and $(\varepsilon_{\mathbf{k}}^2 - k_y^2)< \left|\Delta\right|$, all eigenstates are propagating states with purely real $k_x$. The corresponding group velocity is given as $v_x^{(\lambda\eta)} = \partial \varepsilon_{\mathbf{k}} / \partial k_x^{(\lambda\eta) } = \lambda \eta  \left( \varepsilon_{\mathbf{k}}^2 - k_y^2 \right)^{1/2} / \varepsilon_{\mathbf{k}} $
By comparing $v_x^{(\lambda\eta)}$ with $k_x^{(\lambda\eta)}$, the eigenstate can be determined as electron(hole)-like if the product $\left(\lambda \eta / \varepsilon_{\mathbf{k}}\right)$ has the same (opposite) sign as $\lambda$ which signifies group velocity being (anti)-parallel with the wavevector.  The group velocity and the electron/hole nature of the $(\lambda, \eta)$ branches are shown in Fig. 8. 
This corresponds to the low energy regime in which the energy dispersion splits into two distinct Dirac cones. 
For $(\varepsilon_{\mathbf{k}}^2 - k_y^2)> \left|\Delta\right|$, the two branches of $\eta = -1$ merge and the corresponding eigenstate become evanescent due to the merging of Dirac cones.

The second pseudospinor-component can be solved as
\begin{equation}
\phi_B^{(\lambda\eta)} = \frac{\eta \sqrt{\varepsilon_k^2 - k_y^2} + ik_y}{\varepsilon_k} e^{ik_x^{(\lambda \eta)}x }.
\end{equation}
Correspondingly, the normalized eigenstate outside the barrier and that inside the barrier are given, respectively, as
\begin{subequations}
	\begin{equation}
	\psi^{(\lambda\eta)}(x) = \frac{1}{\sqrt{2}} 
	\begin{pmatrix}
	1 \\
	\frac{\eta \sqrt{\varepsilon_\mathbf{k}^2 - k_y^2} + ik_y}{\varepsilon_\mathbf{k}} 
	\end{pmatrix}
	e^{ik_x^{(\lambda \eta)}x }
	\end{equation}
	\begin{equation}
	\tilde{\psi}^{(\lambda\eta)}(x) = \frac{1}{\sqrt{2}} 
	\begin{pmatrix}
	1 \\
	\frac{\eta \sqrt{\varepsilon_\mathbf{k}^2 - k_y^2} + ik_y}{\varepsilon_\mathbf{k} - V_g} 
	\end{pmatrix}
	e^{iq_x^{(\lambda \eta)}x }
	\end{equation}
\end{subequations}
where $q_x^{\lambda\eta} = \lambda \sqrt{ \eta \left[\left( \varepsilon_k - U_0\right) ^2 - k_y^2\right]^{1/2} - \Delta }$. 
The pseudospin vector, $\mathbf{S} = (S_x, S_y)$, can be determined as
\begin{equation}
\mathbf{S} = \psi^{(\lambda\eta) \dagger} \bm{\sigma} \psi^{(\lambda\eta)} = \left(\frac{k_x^2 + \Delta  }{\varepsilon_{\mathbf{k}}},   \frac{k_y}{\varepsilon_{\mathbf{k}}} \right).
\end{equation}
In Fig. 10, the $x$- and $y$-components of the pseudospin is plotted with $\Delta = -1$. 
At the vicinity of the Dirac points $\mathbf{k} = (\pm \sqrt{ \left|\Delta\right|}, 0)$, $S_x$ and $S_y$ behaves in a contrasting way.
The $S_y$-component is identical for both valleys while the $S_x$-component exhibits a sign change between the two valleys. 
Thus, $D+$ and $D-$ valleys are indistinguishable for $k_y$-directional transport while a strong valley-contrast manifests in the $k_x$-directional transport and the two valleys have opposite chirality.

\begin{figure} [b]
	\includegraphics[scale= .65]{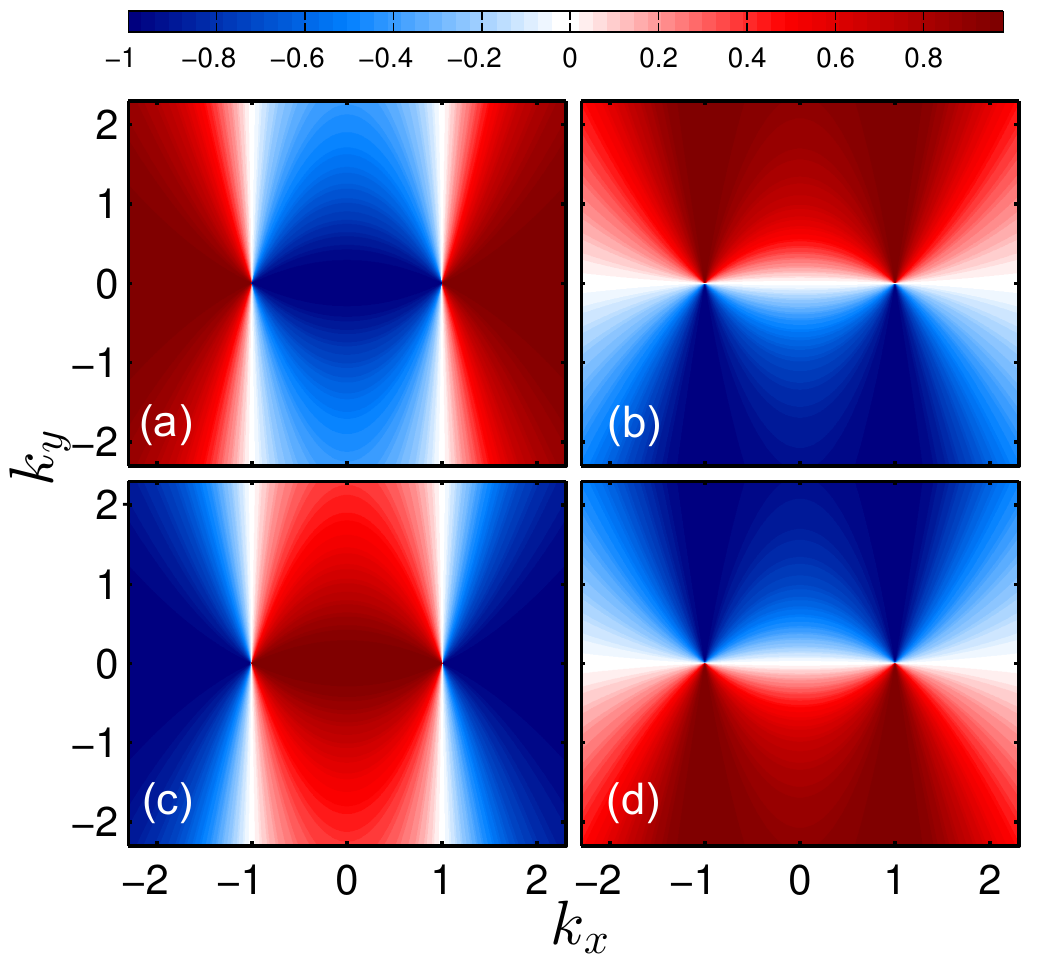}
	\caption{Pseudospin texture of a merging Dirac cone system. (a) and (b) shows, respectively the $x$-component pseudospin and $y$-component pseudospin textures of a merging Dirac cone system for $\varepsilon_{\mathbf{k}} < 0$. (c) and (d), same as (a) and (b) but for $\varepsilon_{\mathbf{k}} > 0$.}
\end{figure}

The total wavefunctions in Region I, II and III are given as
\begin{subequations}
	\begin{equation}
	\Psi_{\text{I}}(x) = \psi^{(\mathcal{D}\mathcal{D})}(x) + \sum_{\eta=\pm} r_{\mathcal{D}}^{(\eta)}\psi^{(-\eta \eta)}(x)
	\end{equation}
	\begin{equation}
	\Psi_{\text{II}}(x) = \sum_{\eta=\pm} a^{(-\eta \eta)}\tilde{\psi}^{(-\eta \eta)}(x) + b^{(\eta\eta)}\tilde{\psi}^{(\eta \eta)}(x),
	\end{equation}
	\begin{equation}
	\Psi_{\text{III}}(x) = \sum_{\eta=\pm} t_{\mathcal{D}}^{(\eta)}\psi^{(\eta \eta)}(x)
	\end{equation}
\end{subequations}
where $t_{\mathcal{D}}^{(\mathcal{D}')}$ and $r_{\mathcal{D}}^{(\mathcal{D}')}$ is the transmission and reflection, respectively. The index $\mathcal{D}, \mathcal{D}' = \pm 1$ denotes the two valleys. 
The transmission and reflection coefficients can be solved by matching $\Psi_{\text{I}}$, $\Psi_{\text{II}}$ and $\Psi_{\text{III}}$ at the boundaries of $x=0$ and $x=d$ via 
\begin{subequations}\label{bc1}
	\begin{equation}
	\Psi_{\text{I}}(x = 0)  = \Psi_{\text{II}}(x=0)  
	\end{equation}
	\begin{equation}
	\Psi_{\text{II}}(x=d) = \Psi_{\text{III}}(x=d)
	\end{equation}
\end{subequations}
and
\begin{subequations}\label{bc2}
	\begin{equation}
	\left. \frac{d\Psi_{\text{I}}(x) }{dx} \right|_{x = 0} = \left. \frac{d\Psi_{\text{II}}(x) }{dx} \right|_{x = 0} 
	\end{equation}
	\begin{equation}
	\left. \frac{d\Psi_{\text{II}}(x) }{dx} \right|_{x = d} = \left. \frac{d\Psi_{\text{III}}(x) }{dx} \right|_{x = d} 
	\end{equation}
\end{subequations}
\begin{widetext}
	The forms a system of equation given as
	\begin{equation}
	\begin{pmatrix}
	\tilde{\mathbb{K}}(0) & -\tilde{\mathbb{Q}}(0) & -\mathbb{Q}(0) & \mathbf{O}_{4\times 2} \\
	\mathbf{O}_{4\times 2}  & -\tilde{\mathbb{Q}}(d) & -\mathbb{Q}(d) & \mathbb{K}(d)
	\end{pmatrix}
	\begin{pmatrix}
	\mathbf{R}^{(\mathcal{D})} \\
	\mathbf{A}^{(\mathcal{D})} \\
	\mathbf{B}^{(\mathcal{D})} \\
	\mathbf{T}^{(\mathcal{D})}
	\end{pmatrix}
	= 
	\begin{pmatrix}
	- \mathcal{K}^{(\mathcal{D})}0) \\
	\mathbf{O}_{4\times 1}
	\end{pmatrix}
	\end{equation}
	where $\mathbf{O}_{M\times N}$ is a $M\times N$ zero matrix, the $4\times2$ matrices, $\mathbb{K}$, $\tilde{\mathbb{K}}$, $\mathbb{Q}$ and $\tilde{\mathbb{Q}}$, are defined as
	\begin{subequations}
		\begin{equation}
		\mathbb{K}(x) \equiv 
		\begin{pmatrix}
		\psi^{(++)}(x) & \psi^{(--)}(x) \\
		k^{(++)}_x \psi^{(++)}(x) & k^{(--)}_x \psi^{(--)}(x)
		\end{pmatrix}
		\text{ , }
		\tilde{\mathbb{K}}(x) \equiv 
		\begin{pmatrix}
		\psi^{(-+)}(x) & \psi^{(+-)}(x) \\
		k^{(-+)}_x \psi^{(-+)}(x) & k^{(+-)}_x \psi^{(+-)}(x)
		\end{pmatrix}
		,
		\end{equation}
		\begin{equation}
		\mathbb{Q}(x) \equiv 
		\begin{pmatrix}
		\tilde{\psi}^{(-+)}(x) & \tilde{\psi^{(+-)}}(x) \\
		q^{(-+)}_x \tilde{\psi}^{(-+)}(x) & q^{(+-)}_x \tilde{\psi}^{(+-)}(x)
		\end{pmatrix}
		\text{ , }
		\tilde{\mathbb{Q}}(x) \equiv 
		\begin{pmatrix}
		\tilde{\psi}^{(++)}(x) & \tilde{\psi}^{(--)}(x) \\
		q^{(++)}_x \tilde{\psi}^{(++)}(x) & q^{(--)}_x \tilde{\psi}^{(--)}(x)
		\end{pmatrix}.
		\end{equation}
	\end{subequations}
	and $\mathcal{K}^{(\mathcal{D})}(x) \equiv \left( \psi^{(\mathcal{D}\mathcal{D})}(x), k_x^{(\mathcal{D}\mathcal{D}))}\psi^{(\mathcal{D}\mathcal{D})}(x) \right)^\mathcal{T}$ with $\mathcal{D} = \pm 1$ indicates the valley index of the incident electron. The transport coefficients are compactly contained in
	\begin{equation}
	\mathbf{R}^{(\mathcal{D})} =
	\begin{pmatrix}
	r_{\mathcal{D}}^{(\mathcal{D})} \\
	r_{-\mathcal{D}}^{(\mathcal{D})}
	\end{pmatrix}
	\text{ , }
	\mathbf{T}^{(\mathcal{D})} =
	\begin{pmatrix}
	t_{\mathcal{D}}^{(\mathcal{D})} \\
	t_{-\mathcal{D}}^{(\mathcal{D})}
	\end{pmatrix}
	\text{ , }
	\mathbf{A}^{(\mathcal{D})} =
	\begin{pmatrix}
	a_{\mathcal{D}}^{(\mathcal{D})} \\
	a_{-\mathcal{D}}^{(\mathcal{D})}
	\end{pmatrix}
	\text{ , }
	\mathbf{B}^{(\mathcal{D})} =
	\begin{pmatrix}
	b_{\mathcal{D}}^{(\mathcal{D})} \\
	b_{-\mathcal{D}}^{(\mathcal{D})}
	\end{pmatrix}
	\end{equation}
\end{widetext}
Finally, the probability current conservation can be written as
\begin{equation}
1  = \sum_{\mathcal{D}' = \pm} \left(-\frac{ v_x^{(-\mathcal{D}' \mathcal{D}' )} }{v_x^{(\mathcal{D} \mathcal{D}) } } \left|r_{\mathcal{D}'}^{(\mathcal{D})}\right|^2 + \frac{ v_x^{(\mathcal{D}' \mathcal{D}' )} }{v_x^{(\mathcal{D} \mathcal{D}) } } \left|t_{\mathcal{D}'}^{(\mathcal{D})}\right|^2 \right)
\end{equation}
where the velocity expectation value is defined as
\begin{equation}
v_x^{(uu') } \equiv \psi^{(uu') \dagger }\frac{\partial \hat{\mathcal{H}}_\mathbf{k}}{\partial k_x} \psi^{(uu') }
\end{equation}
The reflection and tunneling probabilities can then be solved as
\begin{equation}
R^{(\mathcal{D})}_{\mathcal{D}'}  = -\frac{ v_x^{(-\mathcal{D}'\mathcal{D}')} }{v_x^{(\mathcal{D} \mathcal{D}) } } \left|r_{\mathcal{D}' }^{(\mathcal{D})}\right|^2
\text{ , }
T^{(\mathcal{D})}_{\mathcal{D}'}  = \frac{ v_x^{(\mathcal{D}'\mathcal{D}')} }{v_x^{(\mathcal{D} \mathcal{D}) } } \left|t_{\mathcal{D}'}^{(\mathcal{D})}\right|^2
\end{equation}
The $\mathcal{D}$-polarized ballistic tunneling current, under bias voltage $V_B$, is given as
\begin{eqnarray}
\mathcal{J}^{(\mathcal{D})} (V_B, T) &=& \frac{4e  \varepsilon_0 k_0}{(2\pi)^2\hbar} \int dk_x dk_y \left( \frac{\partial \varepsilon_{\mathbf{k}}}{\partial k_x} \right) \nonumber \\
&\times& T(k_y, \varepsilon_{\mathbf{k}}, V_g) f(\varepsilon_{\mathbf{k}}, T)
\end{eqnarray}
where $f(\varepsilon_{\mathbf{k}}, T)$ is the Fermi-Dirac distribution function. At low temperature and small bias voltage, $f(\varepsilon_{\mathbf{k}}, T) \to \left( eV/\varepsilon_0 \right) \delta (\varepsilon_{\mathbf{k}} - \varepsilon_{F} ) $. The ballistic conductance, $G^{(\mathcal{D})} = \mathcal{J}^{(\mathcal{D})} / V_B$, becomes 
\begin{equation}
G^{(\mathcal{D})} (\varepsilon_F, V_g) =  G_0 \int dk_y T(k_y, \varepsilon_{F}, V_g)
\end{equation} 
where $G_0 \equiv  W g_0 k_0 / 2\pi$ and $g_0 \equiv 4e^2/h$.

\begin{figure} [b]
	\includegraphics[scale= .75]{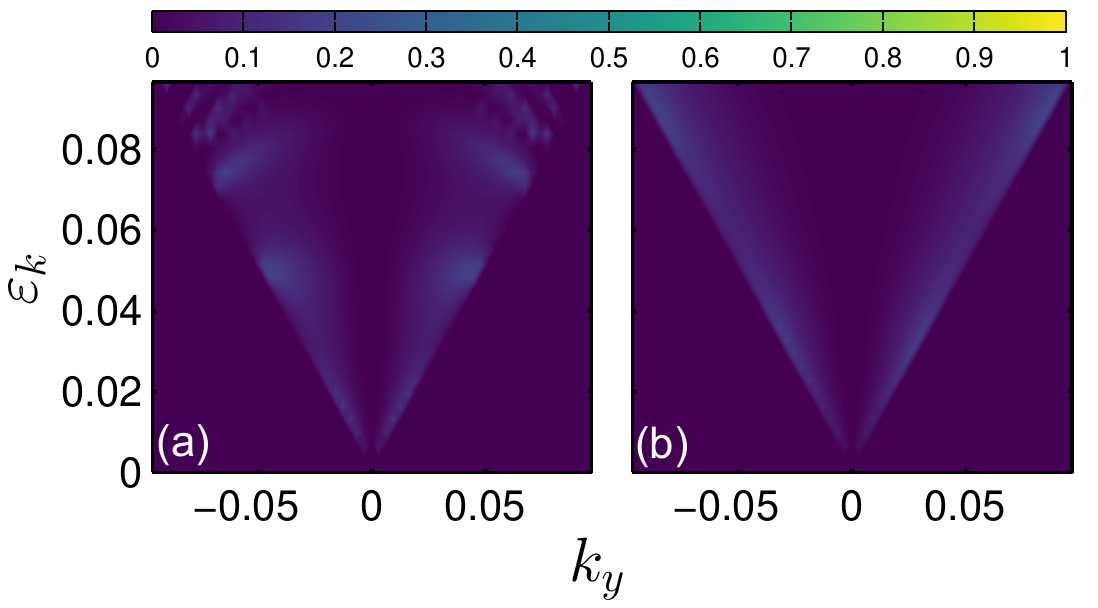}
	\caption{Inter-valley transmission probabilities, $T^{(\pm)}_\mp$, plotted with (a) $V_g = 0.2$ and (b) $V_g = -0.2$ ($d=75$, $\Delta = 0.1$). Note that $T^{(\pm)}_\mp = T^{(\mp)}_\pm$. }
\end{figure}

Before closing this section, we briefly discuss the inter-valley scattering probabilities, $T^{(\pm)}_\mp$, which follows the following symmetry
\begin{equation}
T^{(\pm)}_\mp(\varepsilon_\mathbf{k}, V_g) = T^{(\mp)}_\pm(\varepsilon_\mathbf{k}, V_g)
\end{equation}
The numerical results of $T^{(\pm)}_\mp$ is shown in Fig. 9. It can be seen that due to the mismatching of pseudospin, inter-valley scattering is negligibly small and is only slightly raised at large $|k_y|$ due to the presence of a narrow `strip' of forward propagation branch at large $|k_y|$ that matches the chirality of opposing valleys [see pseudospin winding configuration in Figs. 3(b) and (c)].

\begin{figure} [t]
	\includegraphics[scale= .5]{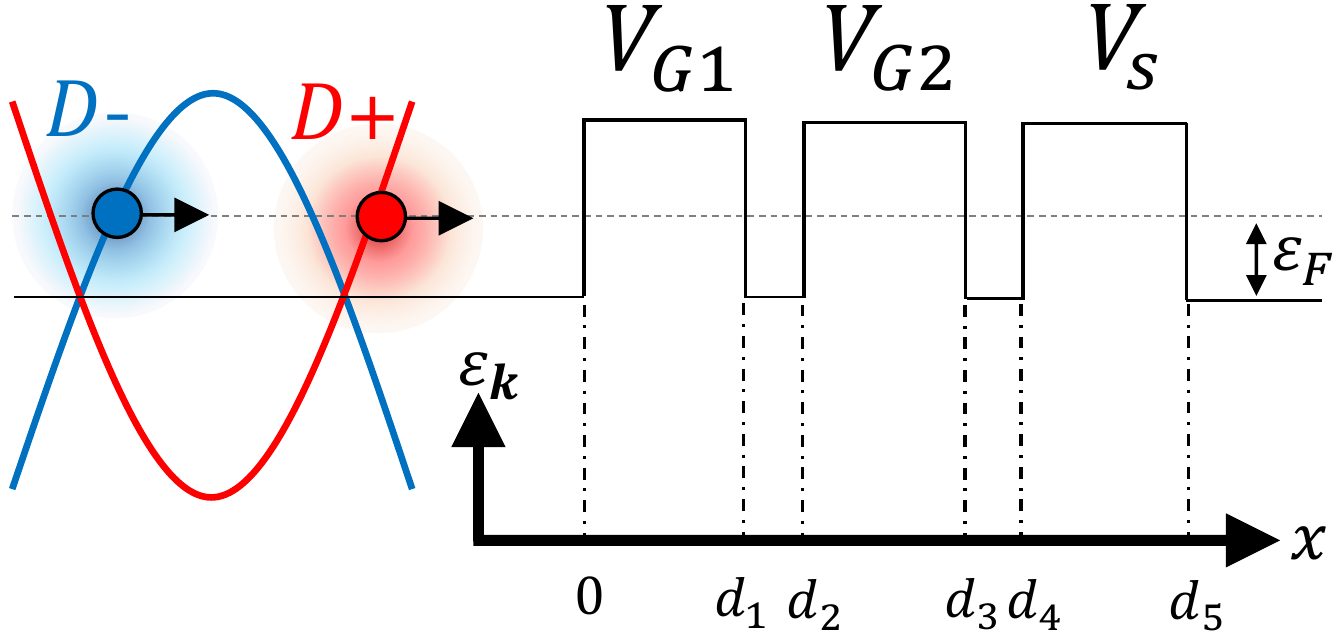}
	\caption{Tunneling band diagram of valley valve device.}
\end{figure}

\section{Modeling of tunneling conductance in a valley valve}
The chiral valley valve is modeled using the following potential profile [see Fig. 10 for the tunneling structure]:
\begin{equation}
U(x) =
\begin{cases}
V_{G1} \text{ , } 0<x<d_1 \\
V_{G2} \text{ , } d_2<x<d_3 \\
V_{s} \text{ , } d_4<x<d_5 \\
0 \text{ , } \text{otherwise}
\end{cases}
\end{equation}
where $V_{G1,2} \equiv e\mathcal{V}_{G1,2} / \varepsilon$ represents the dimensionless form of the first and second gate voltage, $\mathcal{V}_{G1,2}$, and $V_s \equiv e\mathcal{V}_s / \varepsilon_0$ represents the that of the selector voltage, $\mathcal{V}_s$. The barrier widths corresponds to these gates are $d_1$, $(d_3 - d_2)$ and $(d_5 - d_4)$, respectively. Similarly, the transport coefficients can be derived by matching the wavefunctions at each boundary, i.e. $x = 0, d_i$ (where $i = 1,2,4,5$) via Eqs. $(\ref{bc1})$ and $(\ref{bc2})$. This leads to a system of 24 equations which are numerically solved to obtain the transmission coefficients. The corresponding transport probabilities and valley-polarized conductance are calculated via Eqs. (B17) and (B19), respectively. 

\section{Switching characteristics of valleytronic-based NOT-gate}

\begin{figure} [b]
	\includegraphics[scale= .5]{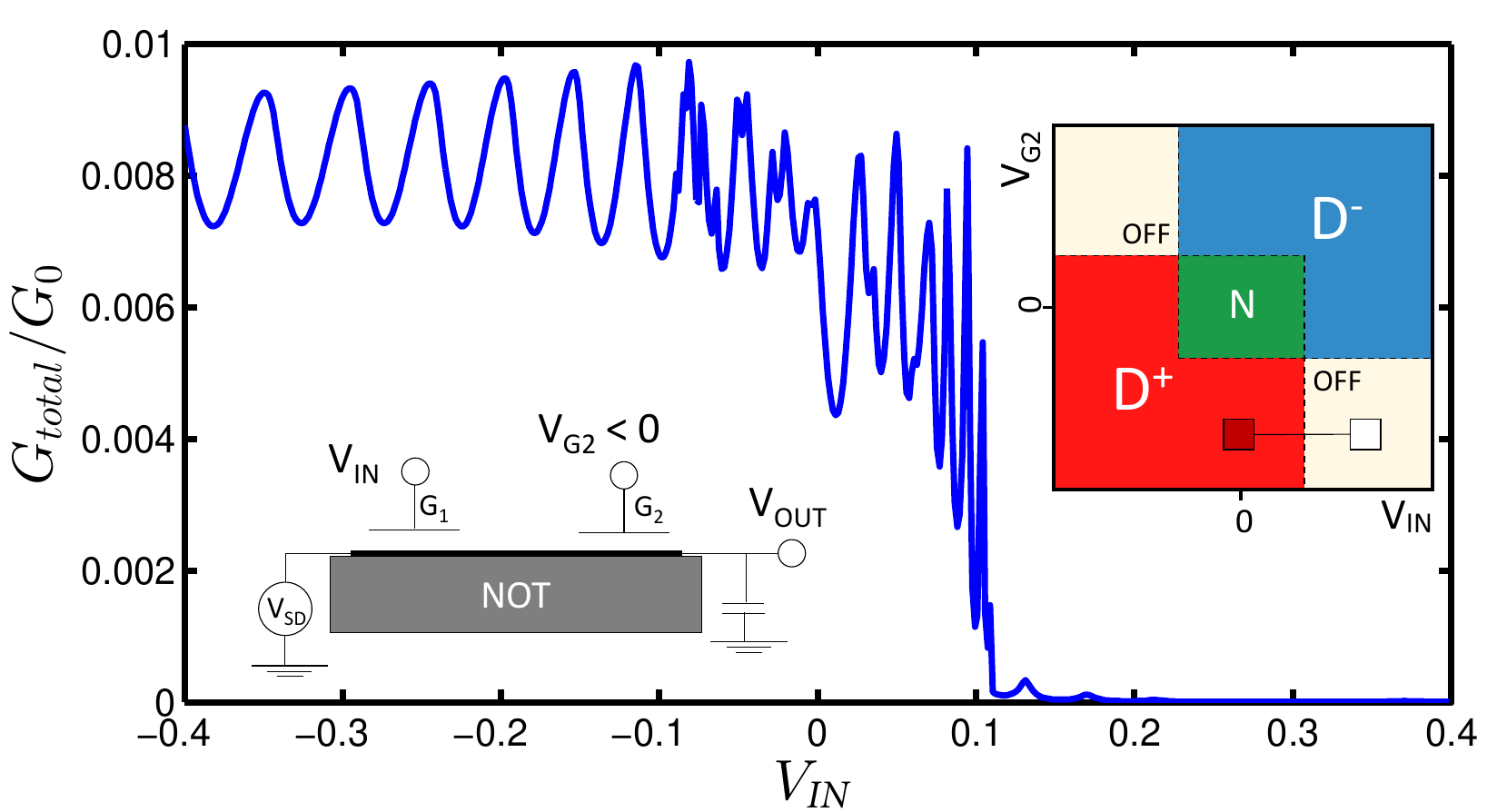}
	\caption{Conductance-voltage characteristic of NOT-gate with a fixed $V_{G2} = 0.2$. Insets show the Boolean line of NOT operation in the VPD with $V_s = 0$, and the gate configurations.}
\end{figure}

As a proof-of-concept, we explicitly show the conductance-voltage characteristics of a valleytronic-based NOT-gate in Fig. 11. The total conductance as a function of input voltage, $V_{\text{IN}}$, is calculated as $G_{\text{total}}(V_{\text{IN} }) = G_+(V_{\text{IN} }) + G_-(V_{\text{IN} }) $ and the $V_{G2} = -0.2$ is a fixed reference voltage. The input signal is fed into the gate via $V_{\text{IN} }$. For $V_{\text{IN}} \leq 0$, the output total conductance is switched into a stable oscillation between $G_{\text{total}}( V_{\text{IN} } \leq 0 ) \approx 0.007G_0$ and $G_{\text{total}} (V_{\text{IN} } \leq 0) \approx 0.009 G_0$ while for $V_{\text{IN}} \geq 0.15$ the output conductance is switched off with $G_{\text{total}}$ in the order of $G_{\text{total}} (V_{\text{IN} } \geq 0.15) \approx 10^{-6}G_0$. This suggests that the implementation of NOT requires a minimum high/low input voltage difference of $\Delta V_{\text{IN}} \approx 0.15 $, which corresponds to a well-achievable value of approximately $150$ meV. The switching time delay can be obtained via a full simulation that takes into account detail device geometry, and is beyond the scope of this work.

\end{appendix}


\begin{thebibliography}{100}
\bibitem{parameter} From first-principle simulation of few-layer black phosphorus with merging Dirac cones \cite{baik}, the average electron effective mass in $x$-direction is $m^* \approx 1.42 m_e$ where $m_e$ is the free electron mass, and the $y$-direction Fermi velocity is $v_F \approx 2.8 \times 10^5$ m/s. Correspondingly, $\varepsilon \equiv 2m^* v_F^2 \approx 1.3$ eV and $k_0 \equiv 2m^*v_F/\hbar \approx 1.1$ nm$^{-1}$.
\end{thebibliography}
\end{document}